\documentclass{article}

\usepackage{arxiv}

\usepackage[utf8]{inputenc} 
\usepackage[T1]{fontenc}    
\usepackage{hyperref}       
\usepackage{url}            
\usepackage{booktabs}       
\usepackage{amsfonts}       
\usepackage{nicefrac}       
\usepackage{microtype}      
\usepackage{lipsum}
\usepackage{graphicx}
\usepackage{amsmath}
\usepackage{float}

\usepackage{amssymb}
\usepackage{algorithm}
\usepackage{algpseudocode}
\usepackage{amsfonts}

\title{Orientation-aware interaction-based deep material network in polycrystalline materials modeling}

\author{
 Ting-Ju Wei \\
  Department of Civil Engineering\\
  National Taiwan University\\
  Taipei, Taiwan \\
   \And
 Tung-Huan Su\thanks{Corresponding author. Email: \texttt{michael.su@ansys.com}} \\
  ANSYS Inc.\\
  Livermore, CA, USA \\
   \And
 Chuin-Shan Chen\thanks{Corresponding author. Email: \texttt{dchen@ntu.edu.tw}} \\
  Department of Civil Engineering\\
  Department of Materials Science and Engineering\\
  National Taiwan University\\
  Taipei, Taiwan \\
}

\begin{document}
\maketitle
\begin{abstract}
Multiscale simulations are indispensable for connecting microstructural features to the macroscopic behavior of polycrystalline materials, but their high computational demands limit their practicality. Deep material networks (DMNs) have been proposed as efficient surrogate models, yet they fall short of capturing texture evolution. To address this limitation, we propose the orientation-aware interaction-based deep material network (ODMN), which incorporates an orientation-aware mechanism and an interaction mechanism grounded in the Hill-Mandel principle. The orientation-aware mechanism learns the crystallographic textures, while the interaction mechanism captures stress-equilibrium directions among representative volume element (RVE) subregions, offering insight into internal microstructural mechanics. Notably, ODMN requires only linear elastic data for training yet generalizes effectively to complex nonlinear and anisotropic responses. Our results show that ODMN accurately predicts both mechanical responses and texture evolution under complex plastic deformation, thus expanding the applicability of DMNs to polycrystalline materials. By balancing computational efficiency with predictive fidelity, ODMN provides a robust framework for multiscale simulations of polycrystalline materials.
\end{abstract}


\section{Introduction}
\label{sec01}

Multiscale simulations commonly utilize representative volume elements (RVEs) to capture the influence of microstructural features on macroscopic responses. However, solving the RVE directly at each material point using finite element (FE) or fast Fourier transform (FFT) solvers can be computationally infeasible for large-scale engineering problems~\cite{temizer2011adaptive, eisenlohr2013spectral, shanthraj2015numerically, vidyasagar2018deformation, lebensohn2020spectral}. To overcome these challenges, machine learning (ML)-based surrogate models, including fully connected neural networks (FCNNs) and recurrent neural networks (RNNs) such as long short-term memory (LSTM) and gated recurrent unit (GRU), have been developed and applied to polycrystal and composite material systems~\cite{li2019machine, zhang2020using, mozaffar2019deep, bishara2023state, frankel2019predicting, chen2021deep}. While these data-driven methods significantly accelerate multiscale simulations, their dependence on extensive retraining for new loading paths or local material law limits their adaptability and applicability.

The deep material network (DMN) provides a micromechanics-based surrogate modeling framework to overcome these limitations~\cite{liu2019deep, liu2019exploring}. Building upon laminate homogenization theory, DMN constructs a hierarchical network of material building blocks to decouple microstructure from local material behaviors. During offline training, DMN learns essential microstructural representations using only linear elastic datasets. Its architecture enables efficient homogenization and de-homogenization during online prediction, achieving accurate nonlinear extrapolation across diverse loading conditions~\cite{wei2023ls, gajek2021fe, su2022multiscale}. This micromechanics foundation grants DMN superior flexibility and predictive accuracy over purely data-driven methods. Furthermore, DMN has been successfully applied to multiscale simulations across various material systems, enhancing computational efficiency and predictive fidelity~\cite{wei2023ls, gajek2021fe, gajek2021efficient}. 

Recent advancements have expanded DMN's capabilities to diverse microstructures and multiphysics problems~\cite{liu2019transfer, huang2022microstructure, jean2024graph, li2024micromechanics, wei2024foundation}. To further enhance computational efficiency and address challenges in modeling multiphase or porous materials, Noels et al.~\cite{noels2022micromechanics, noels2022interaction, wan2024decoding} introduced the interaction-based material network (IMN) as an enhanced variant of DMN. IMN incorporates stress and strain averaging principles and enforces the Hill–Mandel energetic consistency condition. By replacing rotational transformations with stress-equilibrium directions, IMN streamlines the framework while accurately capturing microstructural features.

Despite these advancements, existing DMN and IMN frameworks encounter challenges in modeling polycrystalline materials, particularly in capturing texture evolution under mechanical loading. Crystallographic orientation significantly influences mechanical behavior, highlighting the need for a framework capable of accurately representing and evolving crystallographic texture within an RVE. While IMN's rotation-free material nodes can not capture crystallographic texture, the original DMN struggles to decouple crystallographic orientations from the homogenization function.

To address these challenges, we propose an orientation-aware interaction-based deep material network (ODMN), which builds upon the IMN framework to extend DMN by incorporating an orientation-aware mechanism. This important extension enables ODMN to model mechanical responses and texture evolution in polycrystalline materials with high computational efficiency. The key contributions of ODMN are as follows:
\begin{enumerate}
    \item Orientation-aware mechanism: ODMN introduces an orientation-aware mechanism at the material nodes, enabling the accurate representation of crystallographic orientation distribution within the RVE.

    \item Decoupling of crystallographic orientation and stress-equilibrium direction: ODMN effectively decouples crystallographic orientations from the stress-equilibrium directions within the material network, facilitating accurate texture representation and evolution during deformation.

    \item Integration with crystal plasticity models: During online prediction, learned crystallographic orientations serve as the initial texture and are seamlessly integrated with crystal plasticity models, enabling accurate nonlinear predictions across various loading conditions.

\end{enumerate}

Our results demonstrate that ODMN predicts homogenized mechanical responses and captures texture evolution in polycrystalline material systems. This dual capability addresses longstanding challenges in polycrystalline modeling and significantly advances multiscale simulations.

The remainder of the paper is organized as follows. Section~\ref{sec02} introduces the ODMN architecture and its theoretical foundations. Section~\ref{sec03} details the offline training methodology and dataset generation. Section~\ref{sec04} describes the online prediction framework and its integration with crystal plasticity models. Section~\ref{sec05} presents two case studies on single-phase RVEs and a two-phase polycrystal RVE. Finally, Section~\ref{sec06} summarizes the key findings and concludes the paper.

\section{Orientation-aware interaction-based Deep Material Network}
\label{sec02}

This section presents the architecture of ODMN, detailing its fundamental components and their interactions. The orientation-aware mechanism illustrates how it decouples crystallographic orientations from stress-equilibrium directions, enabling precise texture evolution modeling. Subsequently, the homogenization function is presented, demonstrating how microstructural information is aggregated into effective material properties. Finally, the section summarizes ODMN’s key features and highlights its seamless integration of the orientation-aware and interaction mechanisms as a surrogate model for polycrystalline material behavior.

\subsection{Architecture of the ODMN}

\begin{figure}[htbp]
    \centering
    \includegraphics[width=90mm]{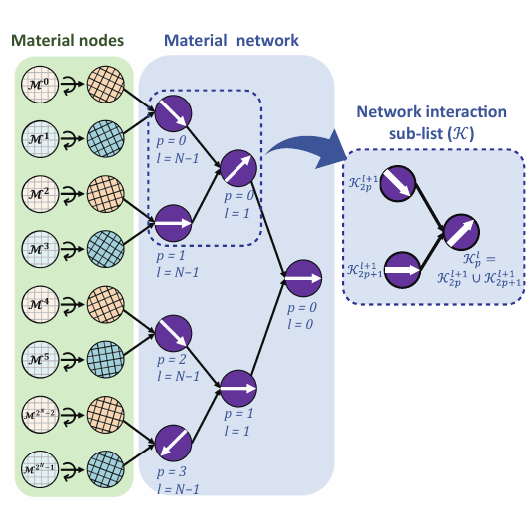}
    \caption{
        Schematic of the ODMN architecture. The material network is structured as a binary tree of depth \( N \). The green area represents the \( 2^N \) material nodes \( \mathcal{M}^{i} \), each responsible for learning the phase volume fractions and crystallographic orientation distributions. The purple circles denote the interaction mechanisms that enforce the stress-equilibrium directions \( \vec{\mathbf{N}}^{l}_{p} \) between material nodes. Here, \( l \) and \( p \) indicate the depth and position of a tree node within the material network, respectively.
    }
    \label{fig: ODMN architecture}
\end{figure}

The ODMN framework consists of a material network structured as a binary tree of depth \( N \) and a set of material nodes, as illustrated in Fig.~\ref{fig: ODMN architecture}. This architecture is specifically designed to capture the microstructural characteristics of polycrystalline materials by integrating two critical physical mechanisms: the \textit{interaction mechanism}, which enforces stress-equilibrium directions within the material network, and the \textit{orientation-aware mechanism}, which represents the crystallographic texture of the RVE.

\newpage

\subsubsection{Material nodes}
In the ODMN, each of the \( 2^N \) material nodes \( \mathcal{M}^{i} \) represents a distinct subregion of the RVE. These material nodes are parameterized by trainable weights, designed to learn the crystallographic orientation distribution and phase volume fraction, thereby modeling the local mechanical behavior of their respective subregions. Specifically:

\begin{itemize}
    \item \textbf{Orientation angles} \( \alpha^i \), \( \beta^i \), and \( \gamma^i \): These Tait–Bryan angles define the crystallographic orientation of the material node by transforming the crystal coordinate system to the specimen coordinate system through sequential rotations:
    \begin{itemize}
        \item \( \alpha^i \): rotation about the \( x \)-axis,
        \item \( \beta^i \): rotation about the \( y \)-axis,
        \item \( \gamma^i \): rotation about the \( z \)-axis.
    \end{itemize}
    The rotations are applied in the order \( \alpha^i \rightarrow \beta^i \rightarrow \gamma^i \).
    
    \item \textbf{Activation parameter} \( z^i \): Each scalar parameter $z^i$ is processed through the softplus activation function to compute the weighting factor $W^i$:
    \begin{equation}\label{eq:activation_function}
        W^i = \operatorname{softplus}(z^i) = \ln\left(1 + \exp\left(z^i\right)\right)
    \end{equation}
    The weighting factor \( W^i \) quantifies the material node's contribution to the overall RVE mechanical behavior and influences the representation of the crystallographic texture.
\end{itemize}

By optimizing these parameters during training, ODMN refines the representation of each subregion, ensuring an accurate and efficient characterization of the microstructure.

\subsubsection{Material network}
The material network, represented by the blue area of Fig.~\ref{fig: ODMN architecture}, is structured as a binary tree. Each tree node corresponds to an individual interaction mechanism designed to enforce stress equilibrium between subregions of the RVE by satisfying the Hill-Mandel condition, as described in the IMN framework~\cite{noels2022micromechanics,noels2022interaction}. The material network is defined by two key components:

\begin{itemize}
    \item \textbf{Sub-list of material nodes $\mathcal{K}^{l}_{p}$}: The sub-list identifies material nodes that constitute the subregion of the RVE at tree level \( l \) and position \( p \). The interaction mechanisms are recursively constructed from the bottom up, progressively aggregating smaller subregions into larger ones:

    \begin{equation}\label{eq:sublist_defined}
        \begin{cases}
            \mathcal{K}^{l}_{p} = \mathcal{K}^{l+1}_{2p} \cup \mathcal{K}^{l+1}_{2p+1}, & \text{if } l < N-1, \\[10pt]
            \mathcal{K}^{l}_{p} = [\mathcal{M}^{2p},\, \mathcal{M}^{2p+1}], & \text{if } l = N-1.
        \end{cases}
    \end{equation}
    At the deepest level (\( l = N - 1 \)), each interaction mechanism involves two material nodes \(\mathcal{M}^{2p}\) and \(\mathcal{M}^{2p+1}\), representing the smallest subregions of the RVE. For \( l < N - 1 \), interaction mechanisms recursively merge child nodes, progressively forming larger regions until the entire RVE is represented at the root node.
    
    \item \textbf{Stress-equilibrium direction} \(\vec{\mathbf{N}}^{l}_{p}\): This unit vector, originally referred to as the interaction direction in the IMN framework, enforces stress equilibrium between subregions. It is parameterized by two trainable angles, \( \theta^{l}_{p} \) and \( \phi^{l}_{p} \), ensuring that \( \vec{\mathbf{N}}^{l}_{p} \) is remains normalized in $\mathbb{R}^3$:
    \begin{equation}\label{eq:direction_vector}
        \vec{\mathbf{N}}^{l}_{p} = \begin{bmatrix}
            \cos(2\pi\phi^{l}_{p}) \sin(\pi \theta^{l}_{p}) \\
            \sin(2\pi\phi^{l}_{p}) \sin(\pi \theta^{l}_{p}) \\
            \cos(\pi \theta^{l}_{p})
        \end{bmatrix}
    \end{equation}
    By optimizing \( \vec{\mathbf{N}}^{l}_{p} \) during offline training, the material network effectively captures mechanical interactions among subregions while maintaining consistency with the stress equilibrium condition.
\end{itemize}

The trainable parameters in the ODMN framework govern both the orientation-aware mechanism and the interaction mechanism, enabling the model to capture intrinsic microstructural characteristics while ensuring stress equilibrium within the RVE. These parameters are defined as:

\begin{equation}\label{eq:trainable_parameters}
    \begin{aligned}
        \mathcal{F} = & \left\{ z^i, \alpha^i, \beta^i, \gamma^i \mid i = 0, 1, \dots, 2^N - 1 \right\} \\
        & \cup \left\{ \theta^{l}_{p}, \phi^{l}_{p} \mid l = 0, 1, \dots, N - 1;\; p = 0, 1, \dots, 2^l - 1 \right\}
    \end{aligned}
\end{equation}

\subsection{Orientation-aware mechanism}

The orientation-aware mechanism in ODMN explicitly incorporates crystallographic orientations into the material nodes through the trainable parameters \( \alpha^i \), \( \beta^i \), and \( \gamma^i \). During offline training, these rotational transformations are applied at each material node, effectively representing the crystallographic texture of polycrystalline materials.

The mathematical formulation governing the rotation of the stress and strain matrix, along with the corresponding transformation of the stiffness matrix, is detailed as follows. Let \(\boldsymbol{\sigma}\) and \(\boldsymbol{\epsilon}\) denote the Cauchy stress and infinitesimal strain tensors, respectively, expressed in Voigt notation:

\begin{equation}
    \boldsymbol{\sigma}=\left[ \sigma_{11}, \sigma_{22},  \sigma_{33},  \sigma_{23},  \sigma_{13},  \sigma_{12}  \right]^T 
\end{equation}
and 
\begin{equation}
    \boldsymbol{\epsilon}=\left[ \epsilon_{11}, \epsilon_{22},  \epsilon_{33},  2\epsilon_{23},  2\epsilon_{13},  2\epsilon_{12}  \right]^T
\end{equation}

For each material node \(\mathcal{M}^{i}\), the local stress-strain relationship in the crystal coordinate system (subscript $c$) is given by:

\begin{equation}\label{eq:crystal_stress_strain}
    \boldsymbol{\sigma}^{i}_c = \mathbb{C}^i \boldsymbol{\epsilon}^{i}_c,
\end{equation}
where \(\mathbb{C}^i\) is the stiffness matrix in the crystal frame.

When the material node \(\mathcal{M}^{i}\) undergoes a rotation defined by the orientation angles \(\{\alpha^i, \beta^i, \gamma^i\}\), the rotated stress $\boldsymbol{\sigma}^{i}_R$ and strain $\boldsymbol{\epsilon}^{i}_R$ in the specimen (global) coordinate system (subscript $R$) are given by:

\begin{equation}\label{eq:R1}
    \boldsymbol{\sigma}_R^i = \mathbf{Z^{R_1}}(\gamma^i) \mathbf{Y^{R_1}}(\beta^i) \mathbf{X^{R_1}}(\alpha^i)\boldsymbol{\sigma}_c
\end{equation}
and 
\begin{equation}\label{eq:R2}
    \boldsymbol{\epsilon}_R^i = \mathbf{Z^{R_2}}(\gamma^i) \mathbf{Y^{R_2}}(\beta^i) \mathbf{X^{R_2}}(\alpha^i)\boldsymbol{\epsilon}_c
\end{equation}
where $\mathbf{X^{R_1}}$, $\mathbf{Y^{R_1}}$, $\mathbf{Z^{R_1}}$, $\mathbf{X^{R_2}}$, $\mathbf{Y^{R_2}}$, $\mathbf{Z^{R_2}}$ are rotation matrices for stress and strain, respectively. The constitutive relationship in the specimen frame is expressed as:

\begin{equation}
    \boldsymbol{\sigma}^i_R = \mathbb{C}^i_R \boldsymbol{\epsilon}^i_R
\end{equation}
with the rotated stiffness matrix given by:

\begin{equation}\label{eq: stiffness rotation}
    \mathbb{C}^i_R =  \mathbf{Z^{R_1}}(\gamma^i)\mathbf{Y^{R_1}}(\beta^i)\mathbf{X^{R_1}}(\alpha^i)\mathbb{C}^i \mathbf{X^{R_2}}(\alpha^i)^{-1}\mathbf{Y^{R_2}}(\beta^i)^{-1}\mathbf{Z^{R_2}}(\gamma^i)^{-1}
\end{equation}

The explicit forms of these rotation matrices are provided in \ref{appendixA}.

\subsection{Homogenization function}
\begin{figure}[htbp]
    \centering
    \includegraphics[width=90mm]{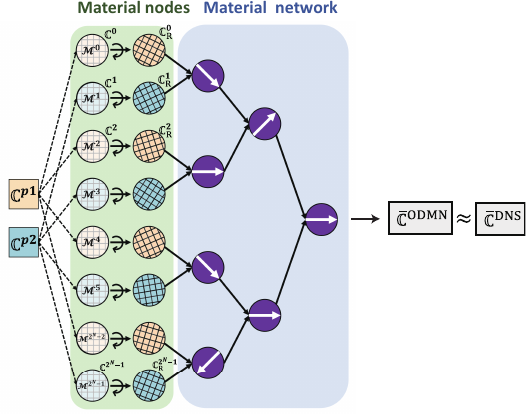}
    \caption{
        Schematic of the ODMN homogenization process. For a two-phase RVE, each material node \( \mathcal{M}^{i} \) is assigned a stiffness matrix \( \mathbb{C}^{i} \) corresponding to its constituent phase (\( \mathbb{C}^{p1} \) or \( \mathbb{C}^{p2} \)). These stiffness matrices are rotated to account for crystallographic orientations, yielding \( \mathbb{C}_{R}^{i} \). The rotated matrices are then recursively aggregated to compute the homogenized stiffness matrix (\( \bar{\mathbb{C}}^{\text{ODMN}} \)).
    }
    \label{fig: homogenization}
\end{figure}

The homogenization function in the ODMN framework integrates the stiffness matrices of individual material nodes into a homogenized stiffness matrix that characterizes the macroscopic mechanical response of the polycrystalline material. As shown in Fig.~\ref{fig: homogenization}, this process involves three key steps:

\begin{enumerate}
    \item \textbf{Stiffness matrix assignment}: Each material node $\mathcal{M}^{i}$ is assigned a stiffness matrix $\mathbb{C}^{i}$ based on its material phase.
    \item \textbf{Rotational transformation}: The assigned stiffness matrix $\mathbb{C}^{i}$ is transformed to $\mathbb{C}_{R}^{i}$ using the orientation angles $\{\alpha^i, \beta^i, \gamma^i\}$, thereby incorporating the crystallographic texture captured by the orientation-aware mechanism.
    \item \textbf{Recursive homogenization}: The rotated stiffness matrices $\mathbb{C}_{R}^{i}$ are recursively aggregated in a bottom-up manner within the binary-tree hierarchy. Interaction mechanisms enforce stress-equilibrium directions and merge child stiffness matrices using the binary homogenization operator $\mathbb{H}_2$.
\end{enumerate}

\newpage

\subsubsection{Stiffness matrix assignment}
Each material node \(\mathcal{M}^{i}\) is assigned a stiffness matrix \(\mathbb{C}^{i}\) based on whether the RVE consists of single-phase or two-phase:

\begin{itemize}
    \item \textbf{For a two-phase RVE}, 

\begin{equation}\label{eq:stiffness_assignment_dualPhase}
     \mathbb{C}^{i} = 
     \begin{cases}
         \mathbb{C}^{p1}, & \text{if } i = 0, 2, \dots, 2^{N} - 2, \\
         \mathbb{C}^{p2}, & \text{if } i = 1, 3, \dots, 2^{N} - 1.
     \end{cases}
\end{equation}
    where \(\mathbb{C}^{p1}\) and \(\mathbb{C}^{p2}\) are the stiffness matrices for the two constituent phases.
    
    \item \textbf{For a single-phase RVE}

\begin{equation}\label{eq:stiffness_assignment_singlePhase}
     \mathbb{C}^{i} = \mathbb{C}^{p1}, \quad \forall i = 0, 1, \dots, 2^{N} - 1.
\end{equation}
where \(\mathbb{C}^{p1}\) represents the stiffness matrix for the single phase.

\end{itemize}

\subsubsection{Rotational transformation}
Each assigned stiffness matrix $\mathbb{C}^{i}$ undergoes a rotational transformation to $\mathbb{C}^{i}_R$, accounting for the crystallographic orientations. This transformation, defined by Eq.~\eqref{eq: stiffness rotation}, is governed by the orientation angles \(\{\alpha^i, \beta^i, \gamma^i\}\) derived from the orientation-aware mechanism. These angles effectively represent the crystallographic texture, capturing the anisotropic characteristics of the RVE.

\subsubsection{Recursive homogenization}

Following rotation, the ODMN recursively homogenizes the stiffness matrices through the material network:

\begin{itemize}
    \item \textbf{Deepest Level(\(l = N-1\))}: At this level, each interaction mechanism merges two rotated stiffness matrices $\mathbb{C}^{2p}_{R}$ and $\mathbb{C}^{2p+1}_{R}$ at position $p$:
    
    \begin{equation}\label{eq:leaf_homogenization}
        \bar{\mathbf{C}}^{N-1}_{p} = \mathbb{H}_2\left(\mathbb{C}^{2p}_{R}, \mathbb{C}^{2p+1}_{R}, \mathcal{M}^{2p},\mathcal{M}^{2p+1}, \vec{\mathbf{N}}^l_p\right)
    \end{equation}

    \item \textbf{Higher Levels (\(l < N-1\))}, The homogenized matrices from child nodes \(\bar{\mathbf{C}}^{l}_{p}\) and \(\bar{\mathbf{C}}^{l+1}_{2p}\) are recursively combined:

    \begin{equation}\label{eq:recursive_homogenization}
        \bar{\mathbf{C}}^{l}_{p} = \mathbb{H}_2\left(\bar{\mathbf{C}}^{l+1}_{2p}, \bar{\mathbf{C}}^{l+1}_{2p+1}, \mathcal{K}^{l+1}_{2p},\mathcal{K}^{l+1}_{2p+1}, \vec{\mathbf{N}}^l_p\right)
    \end{equation}
\end{itemize}

Here, \(\mathbb{H}_2\) represents the binary homogenization operator, formulated based on micromechanical principles~\cite{noels2022interaction}. The detailed mathematical formulation is provided in \ref{appendixB}. This recursive process progresses through the binary tree hierarchy, culminating at the root node (\(l = 0, p = 0\)), where the overall homogenized stiffness matrix is computed as:

\begin{equation}\label{eq:homogenization_function_at_root}
    \mathbb{\bar{C}}^{\text{ODMN}} = \bar{\mathbf{C}}^{0}_{0}
\end{equation}

Notably, no additional rotations are applied beyond those embedded in the material nodes. This design ensures that crystallographic orientations remain decoupled from stress-equilibrium directions, a key feature of ODMN that enables texture evolution while satisfying the Hill–Mandel condition.

\subsection{Summary}
The ODMN captures the microstructural characteristics of polycrystalline materials through two tightly integrated mechanisms:

\begin{enumerate}
    \item \textbf{Interaction mechanism}: Maintains stress equilibrium among RVE subregions. By leveraging stress-equilibrium directions, ODMN encodes mechanical interactions and organizes subregions hierarchically.

    \item \textbf{Orientation-aware mechanism}: Decouples crystallographic orientations from stress-equilibrium directions. Trainable orientation angles enable ODMN to predict texture evolution and anisotropic behavior during deformation.
\end{enumerate}

Within the ODMN framework, material nodes represent RVE subregions, with hierarchical groupings defined through recursive sub-lists. By integrating the interaction mechanism (stress-equilibrium directions) and the orientation-aware mechanism (crystallographic orientations), ODMN provides an efficient and accurate framework for modeling polycrystalline materials.

\section{Offline training }
\label{sec03}
Offline training optimizes the ODMN’s trainable parameters $\mathcal{F}$ to capture the microstructural characteristics of RVE accurately. This training process is achieved by minimizing a relative mean squared error (MSE) objective between the homogenized stiffness predicted by ODMN, $\mathbb{\bar{C}}^{\text{ODMN}}$(see Eq.~\ref{eq:homogenization_function_at_root}) and ground-truth stiffness $\mathbb{\bar{C}}^{\text{DNS}}$ obtained from direct numerical simulations (DNS):

\begin{equation}\label{eq:loss function}
    \text{Loss}(\mathcal{F}) =  \frac{1}{N_{batch}} \sum_{i=1}^{N_{batch}}\frac{\left \| \mathbb{\bar{C}}^{\text{DNS}}_i - \mathbb{\bar{C}}_i^{\text{ODMN}} \right \|^2  }{\left \| \mathbb{\bar{C}}^\text{DNS}_i \right \|^2}
\end{equation}
where $N_{\text{batch}}$ denotes the minibatch size used during training, and $\left\| \cdot  \right\|$ represents the Frobenius norm. The AdamW optimizer~\cite{loshchilov2017decoupled} in PyTorch~\cite{paszke2019pytorch} is used for 200 epochs with a learning rate of 0.001 and minibatch size of 20 for all cases in this study.

To generate ground-truth data, an RVE is constructed in DREAM.3D~\cite{groeber2014dream}, and 500 homogenized stiffness matrices $\mathbb{\bar{C}}^{\text{DNS}}$ are computed using the DAMASK-FFT solver~\cite{roters2019damask}. Among these 500 samples, 400 are allocated for training, while the remaining 100 are reserved for validation.

The ODMN is designed for both single-phase and two-phase polycrystalline materials. The underlying crystal stiffness matrix $\mathbb{C}^{\text{crystal}}$ is parameterized by the elastic constants $\{C_{11}, C_{12}, C_{44}\}$, representing a cubic stiffness matrix in Voigt notation:

\begin{equation}\label{eq:crystal_stiffness_matrix}
\mathbb{C}^{\text{crystal}} =\begin{bmatrix}
C_{11} & C_{12} & C_{12} & 0 & 0 & 0 \\
C_{12} & C_{11} & C_{12} & 0 & 0 & 0 \\
C_{12} & C_{12} & C_{11} & 0 & 0 & 0 \\
0 & 0 & 0 & C_{44} & 0 & 0 \\
0 & 0 & 0 & 0 & C_{44} & 0 \\
0 & 0 & 0 & 0 & 0 & C_{44}
\end{bmatrix}
\end{equation}

To ensure mechanical stability while encompassing diverse microstructural geometries, the elastic constants are sampled from a uniform distribution constrained by the Born stability criteria~\cite{born1940mathematical, mouhat2014necessary}:

\begin{align} \label{eq:stiffness_sampling_space}
C_{11}, C_{12}, C_{44} \sim U[10^{-3}, 10^{3}] \, \text{GPa}, \quad \text{subject to } C_{11} - C_{12} > 0
\end{align}
where, $U(a,b)$ denotes a uniform distribution over $[a,b]$, and $C_{11} - C_{12} > 0$ ensures mechanical stability of the crystal.

For single-phase materials, all material nodes in the ODMN are assigned the same stiffness matrix $\mathbb{C}^{\text{crystal}}$. Elastic constants $\{C_{11}, C_{12}, C_{44}\}$ are sampled from the above distribution, allowing ODMN to learn the homogenization of a wide range of single-phase microstructures.

For two-phase materials, the two stiffness matrices $\mathbb{C}^{p1}$ and $\mathbb{C}^{p2}$ represent Phase 1 and Phase 2, respectively. After sampling $\{C_{11}, C_{12}, C_{44}\}$ for Phase 1, a scaling factor $c_1 \sim 10^{U(-1, 1)}$ is applied to the elastic constants of Phase 2 to introduce stiffness contrast:

\begin{equation}
    C_{ij}^{p2} = c_1 \times C_{ij}^{p2, \text{temp}}, \quad \text{for } ij = 11, 12, 44
\end{equation}

This procedure generates a broad spectrum of stiffness contrasts, enhancing dataset diversity. The detailed algorithm for generating two-phase stiffness matrices is provided in Algorithm~\ref{alg:generate_stiffness_two_phase}.

\begin{algorithm}[htbp]
\caption{Generate Stiffness Matrices for Cubic Two-Phase Materials}
\label{alg:generate_stiffness_two_phase}
\begin{algorithmic}[1]
\Require $N_{\text{samples}}$, $C_{11,\text{range}}$, $C_{12,\text{range}}$, $C_{44,\text{range}}$
\Ensure List of stiffness matrices $\mathbb{C}^{p1}$ and $\mathbb{C}^{p2}$ for Phases 1 and 2
\State Initialize \textit{stiffnessList} $\leftarrow$ [ ]
\While{\textbf{length}(\textit{stiffnessList}) $<$ $N_{\text{samples}}$}
    \State Sample $C_{11}^{p1} \sim U(C_{11,\text{range}})$
    \State Sample $C_{12}^{p1} \sim U(C_{12,\text{range}})$
    \State Sample $C_{44}^{p1} \sim U(C_{44,\text{range}})$
    \If{$C_{11}^{p1} - C_{12}^{p1} > 0$} \Comment{Check stability for Phase 1}
        \State Construct $\mathbb{C}^{p1}$ using $C_{11}^{p1}$, $C_{12}^{p1}$, $C_{44}^{p1}$ and Eq.~\eqref{eq:crystal_stiffness_matrix}
        \State Sample $C_{11}^{p2, \text{temp}} \sim U(C_{11,\text{range}})$
        \State Sample $C_{12}^{p2, \text{temp}} \sim U(C_{12,\text{range}})$
        \State Sample $C_{44}^{p2, \text{temp}} \sim U(C_{44,\text{range}})$
        \If{$C_{11}^{p2, \text{temp}} - C_{12}^{p2, \text{temp}} > 0$} \Comment{Check stability for Phase 2}
            \State Sample $c_1 \sim 10^{U(-1, 1)}$
            \State $C_{11}^{p2} \leftarrow c_1 \times C_{11}^{p2, \text{temp}}$
            \State $C_{12}^{p2} \leftarrow c_1 \times C_{12}^{p2, \text{temp}}$
            \State $C_{44}^{p2} \leftarrow c_1 \times C_{44}^{p2, \text{temp}}$
            \State Construct $\mathbb{C}^{p2}$ using $C_{11}^{p2}$, $C_{12}^{p2}$, $C_{44}^{p2}$ and Eq.~\eqref{eq:crystal_stiffness_matrix}
            \State Append $(\mathbb{C}^{p1}, \mathbb{C}^{p2})$ to \textit{stiffnessList}
        \EndIf
    \EndIf
\EndWhile
\State \Return \textit{stiffnessList}
\end{algorithmic}
\end{algorithm}

\section{Online prediction}
\label{sec04}

\begin{figure}[htbp]
    \centering
    \includegraphics[width=1.0\linewidth]{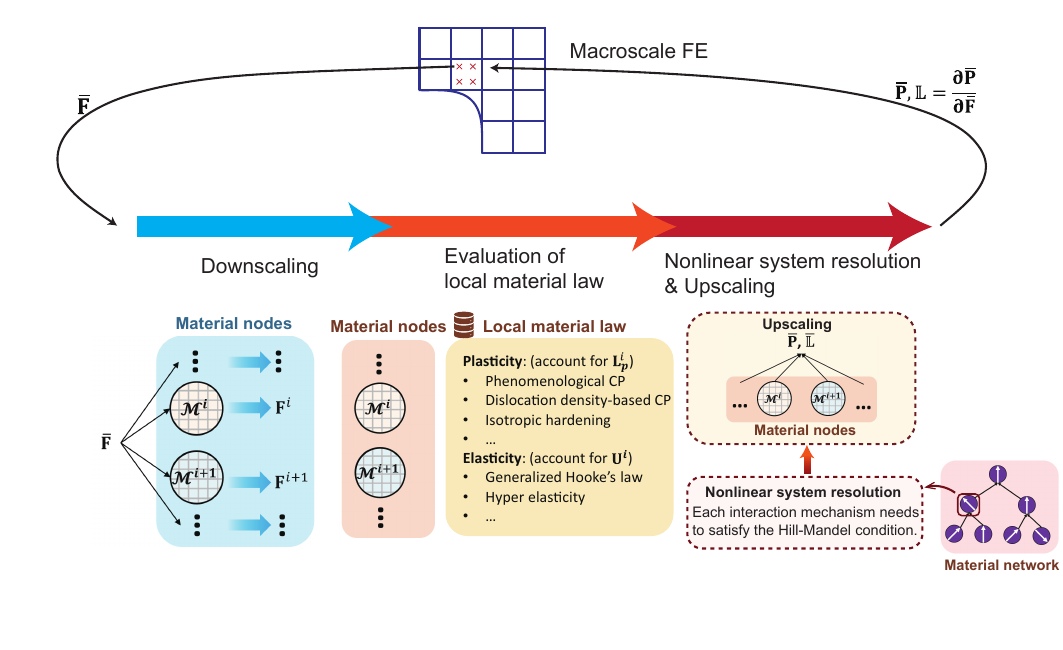}
    \caption{
    Schematic of the ODMN online prediction process, including downscaling the macroscopic deformation gradient $\bar{\mathbf{F}}$, evaluating local material law, resolving the nonlinear system, and upscaling to compute the homogenized stress $\bar{\mathbf{P}}$ and tangent stiffness $\bar{\mathbb{L}}$.
}
    \label{fig:online_prediction}
\end{figure}

During the online prediction phase, the trained ODMN model predicts polycrystalline mechanical behavior and texture evolution under various loading conditions. This process involves key stages: downscaling the macroscopic deformation gradient, evaluating the local material laws, enforcing the Hill–Mandel condition, and upscaling to compute the homogenized response. Figure~\ref{fig:online_prediction} provides an overview of these steps.

\subsection{Downscaling and deformation gradient decomposition}

Downscaling determines the deformation gradient $\mathbf{F}^i$ at each material node $\mathcal{M}^{i}$, linking the macroscopic deformation gradient $\bar{\mathbf{F}}$ to microscale mechanical responses. Using the interaction mapping formalism proposed by Noels et al.~\cite{noels2022interaction}, the deformation gradient at each material node is expressed as:

\begin{equation}\label{eq:interaction_mapping}
   \mathbf{F}^i = \bar{\mathbf{F}} + \sum_{j=0}^{2^{N}-2} \alpha^{i,j} \mathbf{a}^j \otimes \mathbf{N}^j, \quad \text{for } i = 0, 1, \dots, 2^{N}-1 
\end{equation}
where:
\begin{itemize}
    \item $\bar{\mathbf{F}}$: Macroscopic deformation gradient applied to the RVE.

    \item $\mathbf{a}^j$: Interaction variables associated with interaction mechanism $j$ in the material network.

    \item $\mathbf{N}^j$: Stress-equilibrium direction vectors corresponding to interaction mechanism $j$, determined by ODMN parameters.

    \item $\alpha^{i,j}$: Interaction coefficients linking material node $\mathcal{M}^{i}$ to the interaction mechanism $j$, determined by ODMN parameters.
\end{itemize}

The interaction coefficients $\alpha^{i,j}$ are computed based on the weighting factors $W^i$ of the material nodes and the hierarchical structure of the material network. For the $j$-th interaction mechanism at level $l$ and position $p$ within the material network, the associated material nodes are partitioned into two sets: $\mathcal{J}^{j,1}$ and $\mathcal{J}^{j,2}$, representing the sets of material nodes in the first and second child branches, respectively. These sets are determined as follows:

\begin{equation}
    \mathcal{J}^{j,1} = 
\begin{cases}
    \mathcal{K}^{l+1}_{2p}, & \text{if } l < N-1, \\
    \mathcal{M}^{2p}, & \text{if } l = N-1
\end{cases}
\quad
\mathcal{J}^{j,2} = 
\begin{cases}
    \mathcal{K}^{l+1}_{2p+1}, & \text{if } l < N-1, \\
    \mathcal{M}^{2p+1}, & \text{if } l = N-1
\end{cases}
\end{equation}
The interaction coefficients $\alpha^{i,j}$ are defined as:
\begin{equation}
    \alpha^{i,j} = \left\{
    \begin{array}{ll}
        1/ \sum_{i \in \mathcal{J}^{j,1}} W^i, & \text{if } i \in \mathcal{J}^{j,1},\\
        1/ \sum_{i \in \mathcal{J}^{j,2}} W^i, & \text{if } i \in \mathcal{J}^{j,2},\\
        0, & \text{if } i \notin \mathcal{J}^{j,1} \cup \mathcal{J}^{j,2},
    \end{array}
    \right.
\end{equation}

These coefficients ensure that the contributions from all material nodes are appropriately weighted, thereby preserving the Hill–Mandel conditions within the material network.

Once the deformation gradient $\mathbf{F}^i$ at a material node is determined, it is decomposed into elastic and plastic components via the multiplicative decomposition:

\begin{equation} \label{eq: multiplicative decomposition}
     \mathbf{F}^i = \mathbf{F}^i_e \mathbf{F}^i_p
\end{equation}
where $\mathbf{F}^i_e$ is the elastic deformation gradient, representing the recoverable elastic deformation and lattice rotation at the material node $\mathcal{M}^i$, and $\mathbf{F}^i_p$ is the plastic deformation gradient, capturing the irreversible plastic deformation due to crystallographic slip or other plastic mechanisms, such as twinning.

The plastic deformation gradient $\mathbf{F}^i_p$ evolves over time according to the plastic velocity gradient $\mathbf{L}^i_p$:

\begin{equation}
    \dot{\mathbf{F}}^i_p = \mathbf{L}^i_p \mathbf{F}^i_p
\end{equation}

By integrating $\mathbf{L}^i_p$ over time, the plastic deformation gradient $\mathbf{F}^i_p$ is updated accordingly. Consequently, the elastic deformation gradient $\mathbf{F}^i_e$ is obtained from the multiplicative decomposition in Eq.~\eqref{eq: multiplicative decomposition}. The plastic velocity gradient $\mathbf{L}^i_p$ is governed based on the local plasticity material law employed at the material node $\mathcal{M}^i$. 


In summary, the ODMN model provides the flexibility to define local elastic and plasticity material laws at each material node, enabling accurate representation of diverse material behaviors. By updating $\mathbf{F}^i_p$ through the integration of $\mathbf{L}^i_p$ and utilizing $\mathbf{F}^i_e$ to compute the elastic response, the ODMN model can capture the intricate coupling between elastic and plastic deformation mechanisms. Notably, this formulation imposes no restrictions on the choice of material models.

\subsection{Crystallographic orientation initialization and texture evolution}

\begin{figure}[htbp]
    \centering
    \includegraphics[width=90mm]{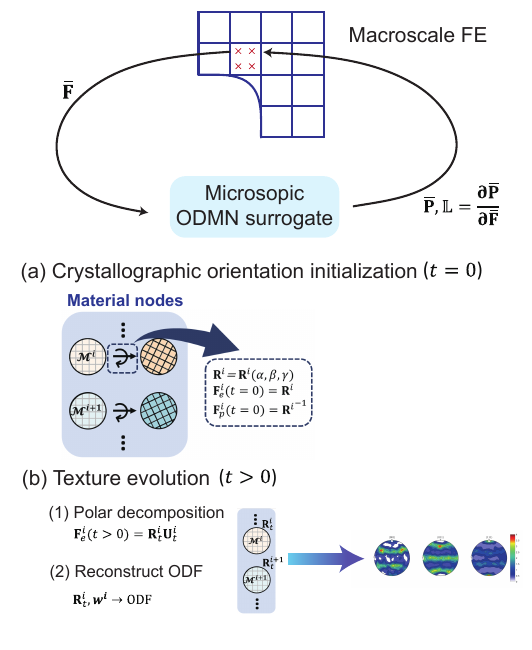}
    \caption{Schematic of (a) crystallographic orientation initialization at $t=0$ and (b) texture evolution during deformation ($t>0$) with updates via polar decomposition and ODF reconstruction.}
    \label{fig:crystal_init_texture_evolution}
\end{figure}

Each material node $\mathcal{M}^{i}$ possesses a distinct crystallographic orientation essential for capturing polycrystalline anisotropic mechanical behavior. The elastic deformation gradient $\mathbf{F}^i_e$ at each material node can be decomposed into a rotation matrix $\mathbf{R}^i$ and a stretch tensor $\mathbf{U}^i$ using polar decomposition:

\begin{equation}
    \mathbf{F}^i_e = \mathbf{R}^i \mathbf{U}^i
\end{equation}

At the initial time step ($t=0$), the elastic and plastic deformation gradients are initialized to align with the initial crystallographic orientation $\mathbf{R}^i$, ensuring a consistent simulation of deformation and texture evolution starting from the initial configuration, as illustrated in Fig.~\ref{fig:crystal_init_texture_evolution}(a):

\begin{equation}\label{eq: FeFp_init}
    \mathbf{F}^i_e(t=0) = \mathbf{R}^i, \quad
    \mathbf{F}^i_p(t=0) = \mathbf{{R}^i}^{-1}
\end{equation}

The rotation matrix $\mathbf{R}^i$ is constructed from the orientation angles $\alpha^i$, $\beta^i$, and $\gamma^i$, which are trainable parameters in the ODMN model. These angles define successive rotations about the coordinate axes:

\begin{equation}
    \mathbf{R}^i =
    \begin{bmatrix}
    1 & 0 & 0 \\
    0 & \cos \alpha^i & -\sin \alpha^i \\
    0 & \sin \alpha^i & \cos \alpha^i \\
    \end{bmatrix}
    \begin{bmatrix}
    \cos \beta^i & 0 & \sin \beta^i \\
    0 & 1 & 0 \\
    -\sin \beta^i & 0 & \cos \beta^i \\
    \end{bmatrix}
    \begin{bmatrix}
    \cos \gamma^i & -\sin \gamma^i & 0 \\
    \sin \gamma^i & \cos \gamma^i & 0 \\
    0 & 0 & 1 \\
    \end{bmatrix}
\end{equation}

During deformation, as illustrated in Fig.~\ref{fig:crystal_init_texture_evolution}(b), the elastic deformation gradient $\mathbf{F}^i_e(t>0)$ is iteratively updated at each loading step until convergence in the macroscale FE simulation. Once converged, the crystallographic orientation $\mathbf{R}^i_t$ is extracted by performing polar decomposition:

\begin{equation} \label{eq:polar decomposition}
    \mathbf{F}^i_e(t>0) = \mathbf{R}^i_t \mathbf{U}^i_t
\end{equation}

The evolution of the rotation matrices $\mathbf{R}^i_t$ characterizes the texture changes in the polycrystalline material during deformation. To reconstruct the orientation distribution function (ODF), the updated orientations $\mathbf{R}^i_t$ are combined with the weighting factors $W^i$ determined from the trained ODMN model. The ODF is computed and visualized in this study using the MTEX software package \cite{bachmann2010texture}.

\subsection{Evaluation of local material law}

The ODMN model provides flexibility in specifying local material behaviors at each material node \( \mathcal{M}^{i} \), enabling the accurate representation of diverse deformation mechanisms in polycrystalline materials. This adaptability is essential for capturing the complex mechanical responses of RVE under various loading conditions.

At each material node \( \mathcal{M}^{i} \), the local mechanical response is characterized by relationships among the deformation gradient \( \mathbf{F}^i(t) \), the first Piola-Kirchhoff stress \( \mathbf{P}^i(t) \), and the internal state variables \( \mathbf{Z}^i(t) \). These relationships are governed by the local material laws \( \mathcal{P}^i \) and \( \mathcal{Z}^i \), which specify the evolution of stress and internal variables as functions of the deformation state:

\begin{equation} \label{eq: localLaw1}
    \mathbf{P}^i(t) = \mathcal{P}^i\big(\mathbf{F}^i(t), \mathbf{Z}^i(t)\big)
\end{equation}

\begin{equation} \label{eq: localLaw2}
    \dot{\mathbf{Z}}^i(t) = \mathcal{Z}^i\big(\mathbf{F}^i(t), \mathbf{Z}^i(t)\big)
\end{equation}

Additionally, the local tangent stiffness is computed by the tangent stiffness operator \( \mathcal{L}^i \), which links incremental changes in the stress to incremental changes in the deformation gradient:

\begin{equation} \label{eq: localLaw3}
    \frac{\partial\mathbf{P}^i}{\partial\mathbf{F}^i}(t) = \mathcal{L}^i\big(\mathbf{F}^i(t),\, \mathbf{Z}^i(t)\big)
\end{equation}

The ODMN framework allows customized definitions of \( \mathcal{P}^i \), \( \mathcal{Z}^i \), and \( \mathcal{L}^i \), accommodating various deformation mechanisms such as dislocation slip, twinning, and phase transformation.



\subsection{Nonlinear system resolution} 
After evaluating the local material responses at each material node, ensuring stress equilibrium at the microscale is essential. In the ODMN framework, this requirement is derived from the Hill–Mandel condition, which guarantees energy consistency between the microscale and macroscale. This approach follows the IMN model proposed by Noels et al.~\cite{noels2022interaction}.

The Hill–Mandel condition leads to stress-equilibrium equations at each interaction mechanism within the ODMN. Specifically, for each interaction mechanism \( j \) (where \( j = 0, 1, \ldots, 2^{N}-2 \)), the stress-equilibrium conditions are given by:

\begin{equation}
    \left(\sum_{i=0}^{2^N-1} W^i \mathbf{P}^i \alpha^{i,j}\right)\cdot \mathbf{N}^j = 0 \text{ for } j=0,...,2^N-2 
\end{equation}
where $W^i$ is the weight associated with material node $\mathcal{M}^{i}$, $\mathbf{P}^i$ represents the first Piola-Kirchhoff stress at material node $\mathcal{M}^{i}$, $\alpha^{i,j}$ denotes the interaction coefficient between material node $\mathcal{M}^{i}$ and $j$-th interaction mechanism, and $\mathbf{N}^j$ is the stress-equilibrium direction associated with the $j$-th interaction mechanism.

The ODMN model enforces the Hill–Mandel condition by minimizing the system residual $\mathbf{r}$, which quantifies the deviation from stress equilibrium at each interaction mechanism. The residual is defined as:

\begin{equation} \label{eq: residuals}
    \mathbf{r} = \sum_{i=0}^{2^N-1} W^i (\mathbf{D}^i)^T \text{vec}(\mathbf{P}^i) 
\end{equation}
where $\text{vec}(\mathbf{P}^i)$ is the vectorized form of the first Piola-Kirchhoff stress tensor at material node $\mathcal{M}^i$, and $\mathbf{D}^i$ is a matrix that encapsulates interaction variables $\mathbf{a}^j$ and stress-equilibrium directions $\mathbf{N}^j$.

The iterative procedure for minimizing the residual $\mathbf{r}$ begins with adjusting the interaction variables $\mathbf{a}^j$, which influence the stress equilibrium conditions. Following each update, the downscaling process is carried out to recompute the deformation gradient at each material node. Once the updated deformation gradient is obtained, the local material laws are evaluated to determine the corresponding stress response. The recalculated stress is then used to compute an updated residual $\mathbf{r}$, which quantifies the deviation from stress equilibrium. If the residual exceeds the predefined convergence threshold, the interaction variables \( \mathbf{a}^j \) undergo further refinement. This iterative process continues until the residual is sufficiently minimized, ensuring the stress equilibrium conditions are satisfied across all interaction mechanisms.

For further details regarding the residual computation and the iterative matrix update process, please refer to ~\ref{appendixD}.

\subsection{Upscaling}
The homogenized stress $\bar{\mathbf{P}}(t)$, which represents the macroscopic stress response, is computed as the weighted average of the local stresses across all material nodes:

\begin{equation}\label{eq:cal_homo_stress}
    \bar{\mathbf{P}}(t) = \frac{1}{\sum_{i=0}^{2^N-1}W^i}\sum_{i=0}^{2^N-1}W^i \mathbf{P}^i
\end{equation}

Similarly, the homogenized tangent stiffness matrix, $\bar{\mathbb{L}}$ is obtained by aggregating the weighted contributions from all material nodes while incorporating the coupling effects between the local stress responses and the global deformation gradient:

\begin{equation}\label{eq:cal_homo_stiffness}
    \text{mat}(\bar{\mathbb{L}}) = \frac{1}{\sum_{i=0}^{2^N-1}W^i}\sum_{i=0}^{2^N-1}W^i \text{mat}(\mathbf{\frac{\partial P}{\partial F}})^i + \mathcal{Y}\frac{\partial \mathbf{A}}{\partial \text{vec}(\mathbf{\bar{F}})}
\end{equation}
where $\mathcal{Y}$ accounts for the coupling between local stress responses and the interaction variables $\mathbf{A}$, and is expressed as:
\begin{equation}
    \mathcal{Y} = \frac{1}{\sum_{i=0}^{2^N-1}W^i} \sum_{i=0}^{2^N-1}W^i \text{mat}(\mathbf{\frac{\partial P}{\partial F}})^i \mathbf{D}^i 
\end{equation}

The operator $\text{mat}(\cdot )$ maps a fourth-order tensor into a matrix. This is done according to the following index mapping rule: 
\begin{equation} \text{mat}(A_{ijkl}) = A_{pq}, \quad \text{where} \ p = i + 3j \ \text{and} \ q = k + 3l \end{equation}

\subsection{Summary}
The overall procedure for online prediction in the ODMN model is summarized in Algorithm~\ref{alg:online_prediction}. The process begins with initializing crystallographic orientations at each material node, ensuring a consistent starting configuration. This is followed by iterative steps that systematically refine the microscale response. First, the macroscopic deformation gradient is downscaled to the material nodes, enabling the evaluation of local material laws. Next, a nonlinear system is solved iteratively to enforce stress equilibrium across interaction mechanisms. Once convergence is achieved, the final step involves upscaling, where the homogenized stress and tangent stiffness matrix are computed to determine the macroscopic material response. This hierarchical approach ensures consistency between microscale deformation mechanisms and macroscale predictions, effectively capturing the anisotropic and nonlinear behavior of polycrystalline materials.

\begin{algorithm}[H]
\caption{ODMN online prediction psudocode}
\label{alg:online_prediction}
\begin{algorithmic}[1]
    \Statex \textbf{Input:} Macroscopic deformation gradient $\bar{\mathbf{F}}$, time increment $dt$, current time $t$
    
    \Statex \textbf{Initialization:}
    \State Initialize iteration counter: $iter \gets 0$ \Comment{iteration counter}
    \If{$t = 0$}
        \State Initialize the deformation gradients at each material node $\mathcal{M}^i$:
        \State \hspace{1em} $\mathbf{F}^i_e(0) \gets \mathbf{R}^i$ \Comment{Elastic deformation gradient, see Eq.~\eqref{eq: FeFp_init}}
        \State \hspace{1em} $\mathbf{F}^i_p(0) \gets \mathbf{R}^{i^{-1}}$ \Comment{Plastic deformation gradient, see Eq.~\eqref{eq: FeFp_init}}
        \State Set interaction variables: $\mathbf{a}^j \gets 0$

    \EndIf

    \While{True}
        \State Increment iteration counter: $iter \gets iter + 1$
        
        \Statex \hspace{1em} Downscale: 
        \State $\mathbf{F}^i \gets \operatorname{downScaling}(\bar{\mathbf{F}})$ 
        \Comment{See Eq.~\eqref{eq:interaction_mapping}}
        
        \Statex \hspace{1em} Evaluate local material response: 
        \State $\mathbf{P}^i, \left(\frac{\partial \mathbf{P}}{\partial \mathbf{F}}\right)^i \gets \operatorname{evalLocalLaws}(\mathbf{F}^i, dt)$
        \Comment{See Eqs.~\eqref{eq: localLaw1}, \eqref{eq: localLaw2}, \eqref{eq: localLaw3}}

        \Statex \hspace{1em} Compute residuals: 
        \State $\mathbf{r} \gets \operatorname{evalResiduals}(\mathbf{P}^i, \mathbf{a}^j, W^i)$
        \Comment{See Eq.~\eqref{eq: residuals}}

        \If{$iter = 1$}
            \State Set initial residual: $r_0 \gets r_{abs}$ 
            
        \EndIf
        
        \State Compute relative residual: $r_{rel} \gets r_{abs} / r_0$
        
        \Statex Check convergence: 
        \State $ \textit{isConvergence} \gets (r_{rel} < tol_{rel}) \ \textbf{or} \ (r_{abs} < tol_{abs})$

        \If{$\text{isConvergence}$}
            \State Compute homogenized stress $\bar{\mathbf{P}}$ 
            \Comment{See Eq.~\eqref{eq:cal_homo_stress}}
            \State Compute homogenized tangent stiffness $\bar{\mathbb{L}}$ 
            \Comment{See Eq.~\eqref{eq:cal_homo_stiffness}}
            \State \textbf{Exit loop}
        \EndIf
         
        \State $\delta \mathbf{A} \gets -\left(\frac{\partial \mathbf{r}}{\partial \mathbf{A}}\right)^{-1} \mathbf{r}$ 
        \Comment{Newton-Raphson update, see Eq.~\ref{eq: deltaA}}

        \State$\mathbf{A} \gets \mathbf{A} + \delta \mathbf{A}$ 
        \Comment{See Eq.~\ref{eq: correctA}}
    \EndWhile

    \Statex Evolution of the crystallographic orientations

    \For{each material node \(\mathcal{M}^i\)}
        \State Update the rotation matrix \(\mathbf{R}^i_t\) \Comment{See Eq.~\eqref{eq:polar decomposition} }
    \EndFor
       
    \State \textbf{Output:} Homogenized stress $\bar{\mathbf{P}}$, homogenized tangent stiffness $\bar{\mathbb{L}}$
\end{algorithmic}
\end{algorithm}

\section{Numerical Results} \label{sec05}

This section presents the numerical results obtained using the ODMN framework. Simulations of single-phase and two-phase polycrystalline RVEs are performed to demonstrate the versatility and predictive capabilities of the ODMN model. The results underscore the model's ability to accurately predict stress-strain behavior and texture evolution under various loading conditions.

\subsection{Single-phase polycrystal}
\begin{figure}[htbp]
\centering
\includegraphics[width=1.0\textwidth]{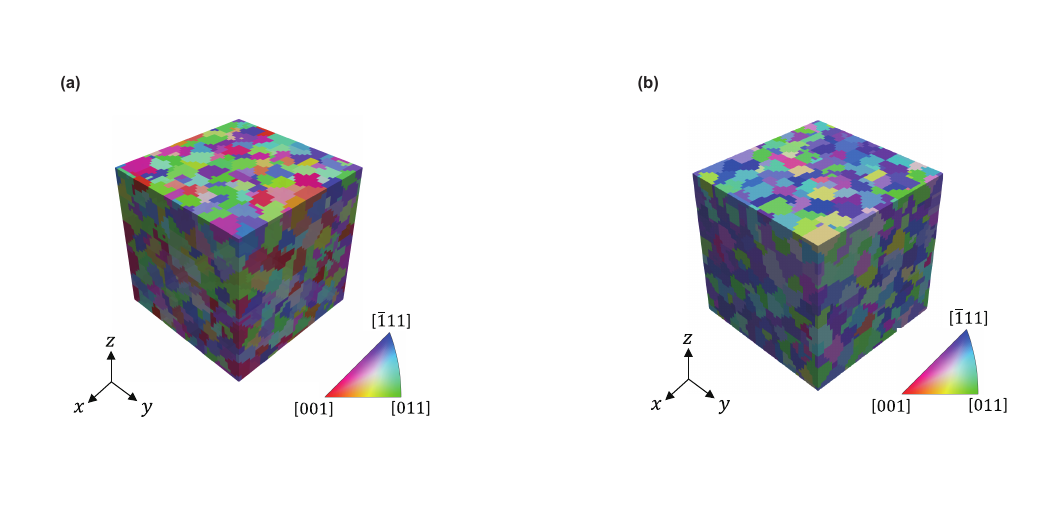}
\caption{
    The geometry of the single-phase polycrystalline RVE used in simulations. (a) RVE with random crystallographic texture. (b) RVE with preferred orientation.
} 
\label{fig: RVE_geometry}
\end{figure}

The single-phase polycrystalline RVEs with equiaxed grains were studied. Two initial textures were considered: one with randomly oriented grains and another with a preferred orientation. Each RVE was discretized on a uniform 45x45x45 grid comprising 809 grains, as shown in Fig.~\ref{fig: RVE_geometry}.  These RVEs were selected to evaluate the robustness and predictive accuracy of the ODMN model under diverse microstructural scenarios.

\newpage

\subsubsection{Offline training results} 

During the offline training, the ODMN model parameters were optimized for random textured and preferred orientation RVEs. The training process performance was evaluated using training curves, illustrating the convergence behavior of the ODMN across its hierarchical layers $N$. An error metric was employed to quantify the discrepancy between the DNS results and the ODMN model predictions for the training and validation datasets. The training/validation error is defined as:

\begin{equation}
\text{Error} = \frac{1}{N_{dataset}} \sum_{i=1}^{N_{dataset}}\frac{\left \| \mathbb{\bar{C}}^{\text{DNS}}_i - \mathbb{\bar{C}}^{\text{ODMN}}(\mathbb{C}^{p1},\mathbb{C}^{p2}) \right \|^2  }{\left \| \mathbb{\bar{C}}^\text{DNS}_i \right \|^2}
\end{equation}

\begin{figure}[htbp]
\centering
\includegraphics[width=1.0\textwidth]{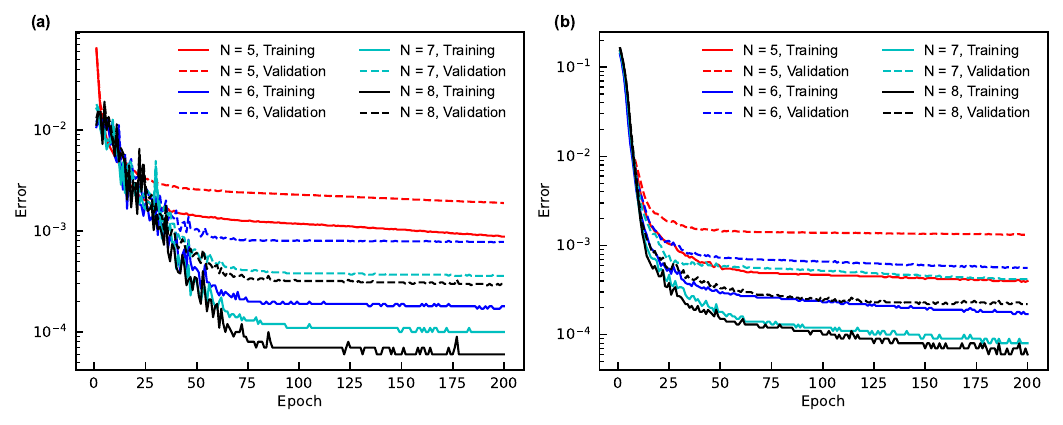}
\caption{
    Training and validation curves for the ODMN model: (a) random textured RVE and (b) preferred orientation RVE across different hierarchical levels $N$.
}
\label{fig:training_curve_singlePhase}
\end{figure}

The training was conducted over 200 epochs for both RVE types. The resulting training curves, presented in Fig.~\ref{fig:training_curve_singlePhase}, demonstrate how increasing the number of hierarchical levels enhances the ODMN learning capacity.

\newpage

\begin{figure}[htbp]
\centering
\includegraphics[width=1.0\textwidth]{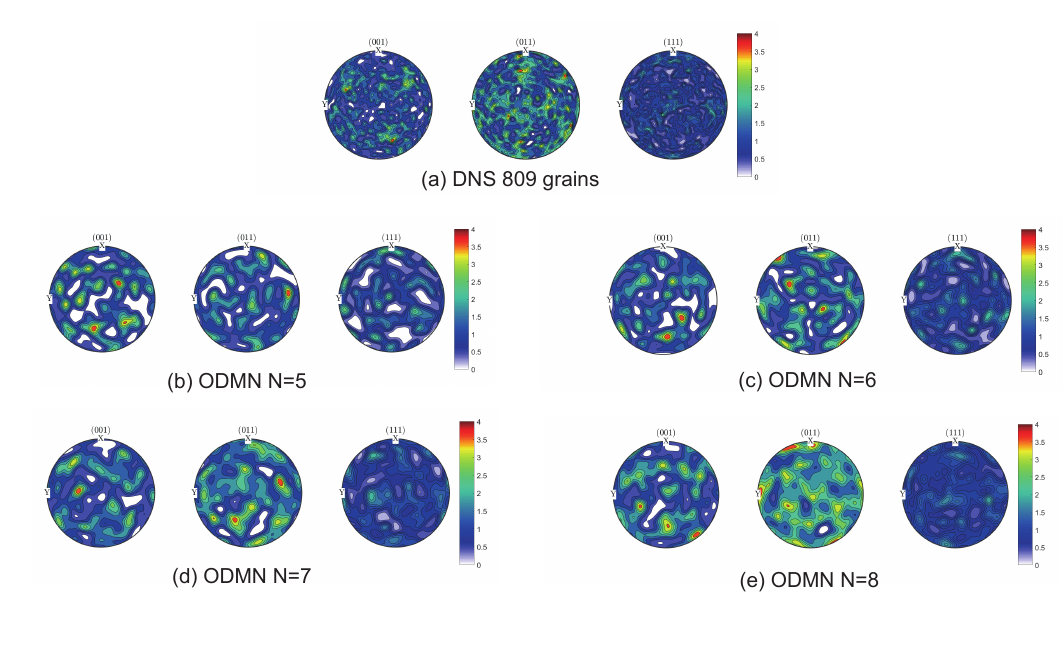}
\caption{
    Pole figure for the random textured RVE after training, comparing DNS results with predictions from the trained ODMN model. 
}
\label{fig:random_pole_figure_after_training}
\end{figure}

\begin{figure}[htbp]
\centering
\includegraphics[width=1.0\textwidth]{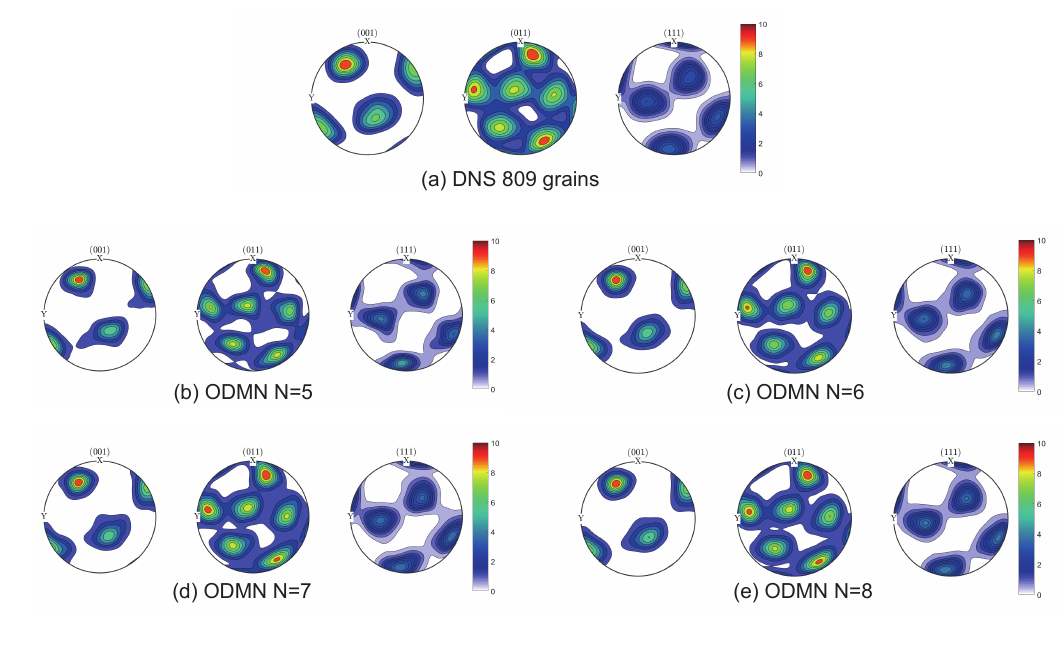}
\caption{
    Pole figure for the preferred orientation RVE after training, comparing DNS results with predictions from the trained ODMN model. 
}
\label{fig:preferred_pole_figure_after_training}
\end{figure}

We visualized the trainable parameters as pole figures to evaluate the ODMN model's capability to capture crystallographic orientation distributions. These figures indicate that the trained ODMN model aligns closely with DNS results, with notable improvements as the number of hierarchical layers increases, as shown in Fig.~\ref{fig:random_pole_figure_after_training} and Fig.~\ref{fig:preferred_pole_figure_after_training}. 

Furthermore, the similarity between the textures predicted by the trained ODMN model and those obtained from DNS was quantitatively assessed using the normalized texture index of the difference ODF(DODF), $\hat{T}^d$. This index quantifies discrepancies between two ODFs and is defined as~\cite{li2005texture}:
\begin{equation}
    \hat{T}^d = \frac{\int [f_{\text{ODMN}}(g)-f_{\text{DNS}}(g)]^2 dg}{\int [f_{\text{DNS}}(g)]^2 dg}
\end{equation}
where $f_{\text{ODMN}}(g)$ represents the ODF predicted by the trained ODMN model, and $f_{\text{DNS}}(g)$ denotes the ODF obtained from DNS.

The results summarized in Table~\ref{tab:texture_index_comparison} show that the normalized texture index $\hat{T}^d$ decreases as the model complexity increases, indicating improved texture capture for random textured and preferred orientation RVEs. This highlights the trained ODMN model's ability to align more closely with DNS results as its architecture becomes more sophisticated.

\begin{table}[htbp]
\centering
\caption{Normalized texture index of the difference ODF ($\hat{T}^d$) between the trained ODMN model and DNS for various ODMN architectures applied to random textured and preferred orientation RVEs.}
\begin{tabular}{lcc}
\hline
ODMN & Random Textured & Preferred Orientation \\
\hline
N = 5 & 0.4755 & 0.1981 \\
N = 6 & 0.2861 & 0.1089 \\
N = 7 & 0.2405 & 0.0879 \\
N = 8 & 0.1179 & 0.0901 \\
\hline
\end{tabular}
\label{tab:texture_index_comparison}
\end{table}

\newpage

\subsubsection{Stress-strain predictions}

During the online prediction phase, the trained ODMN model was evaluated using the phenomenological crystal plasticity model for the plastic behavior and generalized Hooke's law for the elastic response. The material properties at each material node were based on the AA6022-T4 aluminum alloy \cite{barrett2019deep, damask_documentation}. AA6022-T4 features an FCC crystal structure with 12 slip systems, and its material parameters are summarized in Table~\ref{tab:crystal_plasticity_parameters}. The detailed formulation of the phenomenological crystal plasticity model is presented in~\ref{appendixE}, while the elastic behavior is described by the generalized Hooke's law in~\ref{appendixF}.

\begin{table}[htbp]
\centering
\caption{Elastic and plastic material parameters for AA6022-T4 \cite{barrett2019deep, damask_documentation}.}
\label{tab:crystal_plasticity_parameters}
\renewcommand{\arraystretch}{1.2} 
\setlength{\tabcolsep}{10pt} 
\begin{tabular}{cccccccc}
\hline
\textbf{$N_s$} & \textbf{$h_0^{\text{sl-sl}}$}(GPa) & \textbf{$\xi^{\alpha}_{\infty}$}(MPa) & \textbf{$\xi^{0}$}(MPa) & \textbf{$n$} & \textbf{$a$} & \textbf{$\dot{\gamma}_0$}$(\text{s}^{-1})$ & \textbf{$h_{\text{int}}^{\alpha}$} \\
\hline
12 & 1.02 & 266 & 76 & 20 & 3.7 & 0.001 & 0 \\
\hline
\end{tabular}

\vspace{0.5cm}

\begin{tabular}{cccc}
\hline
{$C_{11}$ (GPa)} & {$C_{12}$ (GPa)} & {$C_{44}$ (GPa)} & {$h^{\text{sl-sl}}$} \\
\hline
191 & 162 & 42.2 & [1, 1, 5.123, 0.574, 1.123, 1.123, 1] \\
\hline
\end{tabular}
\end{table}

Two strain rates, $\dot{F}_{11}=0.0001$ and $\dot{F}_{11}=1$, were selected to evaluate the rate dependency of the ODMN model under uniaxial tensile loading conditions. The RVEs were loaded to ${F}_{11}=1.3$. Simulations were performed using both DNS and the ODMN model.

\begin{figure}[htbp]
\centering
\includegraphics[width=1.0\textwidth]{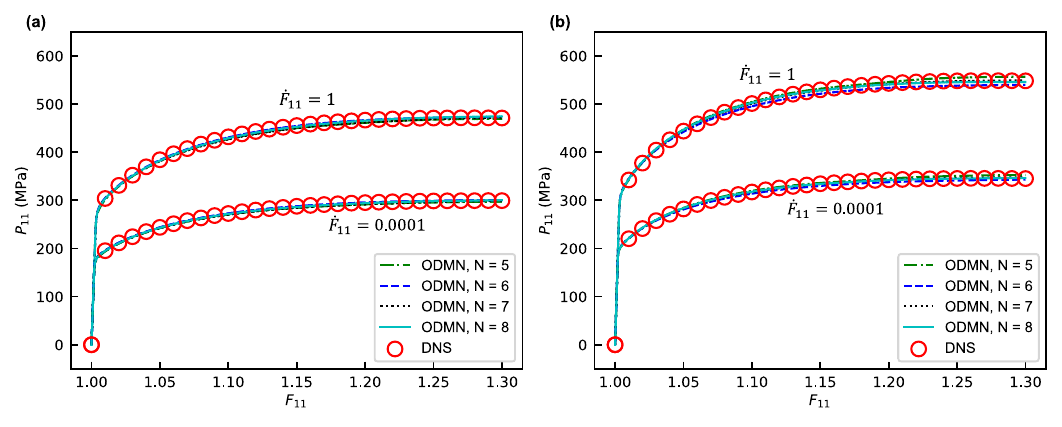}
\caption{
    Stress-strain curves for uniaxial tensile loading of (a) random texture and (b) preferred orientation RVEs at strain rates $\dot{F}_{11}=0.0001$ and $\dot{F}_{11}=1$, comparing DNS results with various ODMN model architectures.
}
\label{fig: ss curve uniaxial tensile}
\end{figure}
\newpage

In addition to uniaxial tensile loading, the ODMN model was evaluated under cyclic and shear loading conditions to assess its predictive capability across different loading scenarios. For cyclic loading, a strain-controlled cycle with a maximum strain of $F_{11} = 1.3$ was applied at a strain rate of $\dot{F}_{11} = 1$. For shear loading, the RVE was subjected to simple shear up to $F_{21} = 0.3$ at a strain rate of $\dot{F}_{21} = 1$.

\begin{figure}[htbp]
\centering
\includegraphics[width=1.0\textwidth]{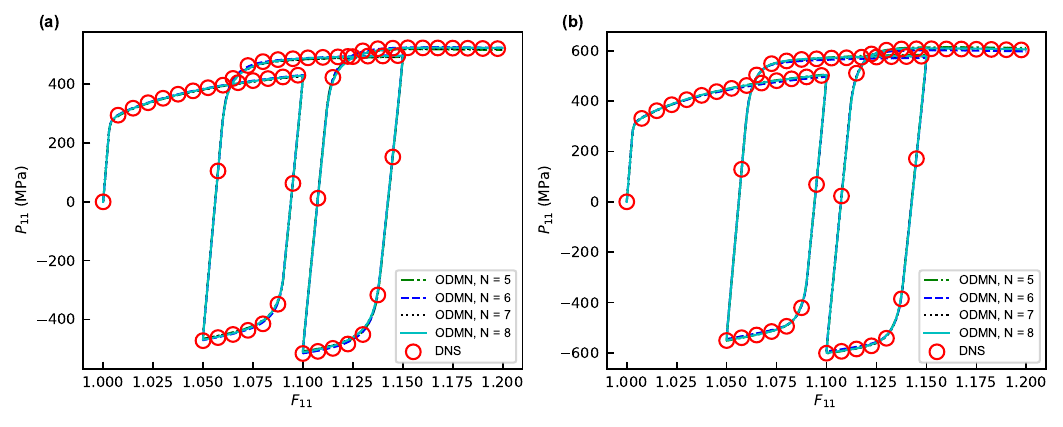}
\caption{
    Stress-strain curves for cyclic loading of (a) random texture and (b) preferred orientation RVEs at a strain rate of $\dot{F}_{11}=1$, comparing DNS results with various ODMN model architectures.
}
\label{fig: ss curve cyclic}
\end{figure}

\begin{figure}[htbp]
\centering
\includegraphics[width=1.0\textwidth]{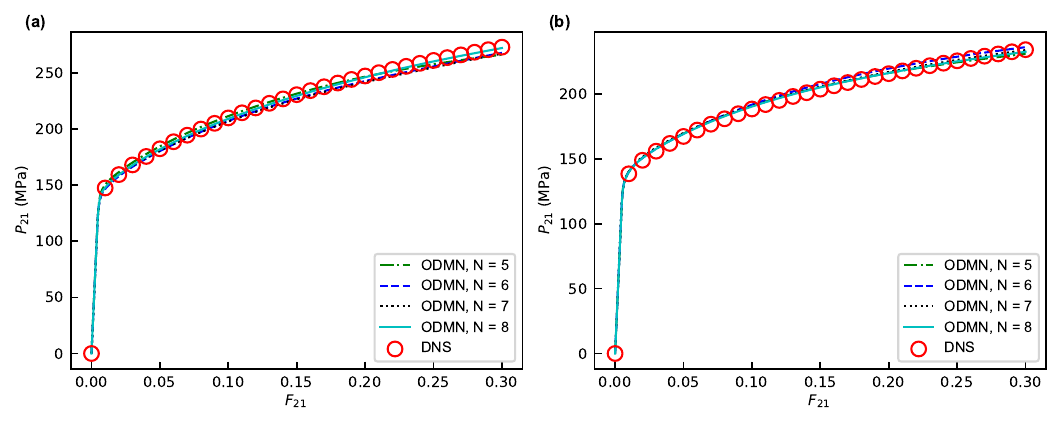}
\caption{
    Stress-strain curves for simple shear loading of (a) random texture and (b) preferred orientation RVEs at a strain rate of $\dot{F}_{21} = 1$, comparing DNS results with various ODMN model architectures.
}
\label{fig: ss curve shear}
\end{figure}

As shown in Fig. \ref{fig: ss curve uniaxial tensile}, Fig. \ref{fig: ss curve cyclic}, and Fig. \ref{fig: ss curve shear}, both the preferred orientation and the random texture RVEs exhibit a high degree of consistency when comparing the results from different hierarchical levels of the ODMN model with those of DNS. Notably, under uniaxial tensile and cyclic loading conditions, even the model architecture with the fewest trainable weights ($N=5$) demonstrates excellent agreement with the DNS results, highlighting the robustness of the ODMN model. However, the model architectures with higher hierarchical levels show superior performance under shear loading, emphasizing the benefit of increased complexity for certain loading scenarios.

\subsubsection{Texture evolution}

\begin{figure}[htbp]
\centering
\includegraphics[width=1.0\textwidth]{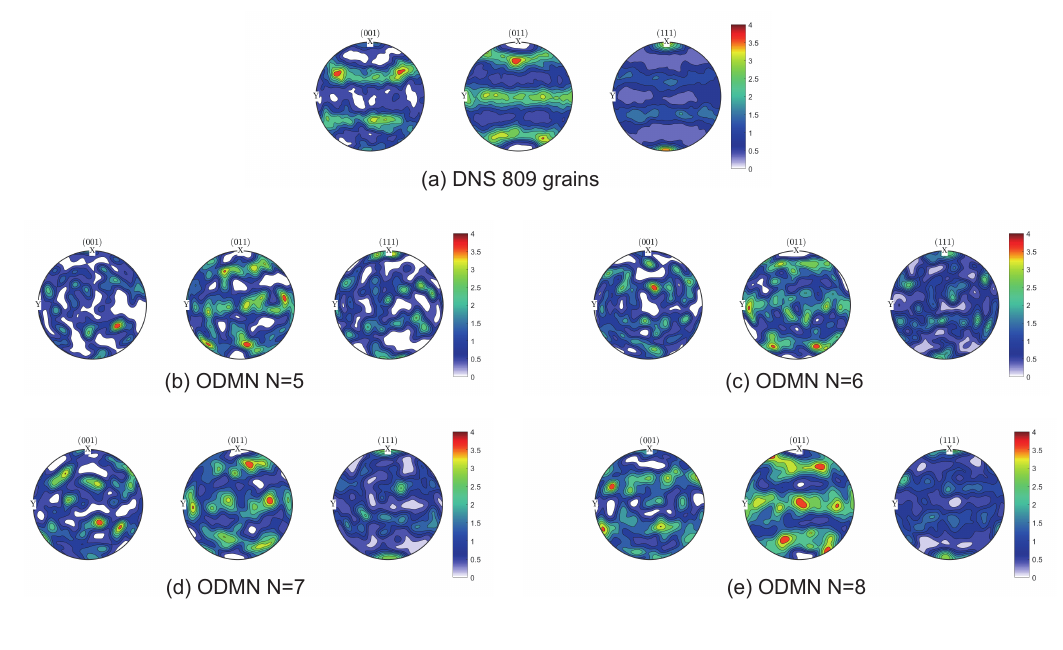}
\caption{
    Pole figures of the random textured RVE after uniaxial tensile loading with an applied deformation of $F_{11}=1.3$ and strain rate $\dot{F}_{11}=1$. DNS results are compared with predictions from the trained ODMN model.
}
\label{fig:random_pole_figure_after_uniaxial}
\end{figure}

\begin{figure}[htbp]
\centering
\includegraphics[width=1.0\textwidth]{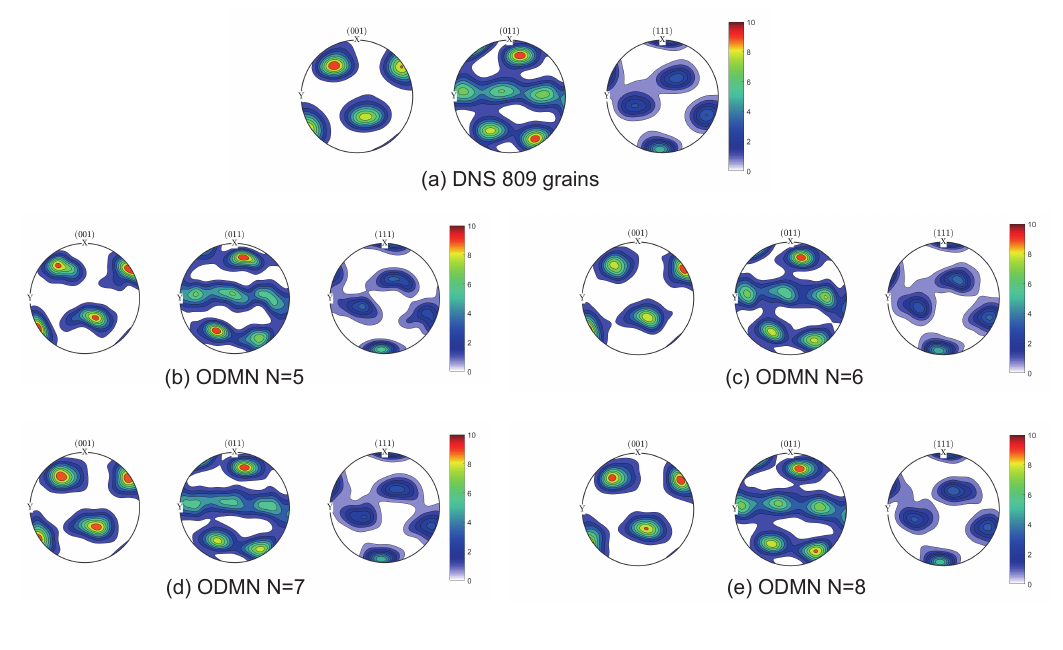}
\caption{
    Pole figures of the preferred orientation RVE after uniaxial tensile loading with an applied deformation of $F_{11}=1.3$ and strain rate $\dot{F}_{11}=1$. DNS results are compared with predictions from the trained ODMN model.
}
\label{fig:preferred_pole_figure_after_uniaxial}
\end{figure}

To further analyze the ODMN performance, the crystallographic textures at ${F}_{11} = 1.3$ were compared for both the preferred orientation and random texture RVEs under a strain rate of $\dot{F}_{11} = 1$. The texture characteristics were evaluated using pole figures, as shown in Fig.~\ref{fig:random_pole_figure_after_uniaxial} and Fig.~\ref{fig:preferred_pole_figure_after_uniaxial}. Additionally, the DODF ($\hat{T}^d$) presented in Table~\ref{tab:texture_index_comparison_uniaxial} provides a quantitative measure of texture evolution. Although the relationship between hierarchical levels $N$ and texture prediction accuracy is not strictly monotonic, the overall trend indicates that increasing the number of hierarchical levels enhances the ODMN model's ability to predict texture evolution.

\begin{table}[htbp]
\centering
\caption{Normalized texture index of the difference ODF ($\hat{T}^d$) between the trained ODMN model and DNS for different model architectures of random textured and preferred orientation RVEs under uniaxial loading with $F_{11}=1.3$ at a strain rate of $\dot{F}_{11}=1$.}
\begin{tabular}{lcc}
\hline
ODMN & Random Textured & Preferred Orientation \\
\hline
N=5 & 0.3557 & 0.1492 \\
N=6 & 0.2210 & 0.0750 \\
N=7 & 0.2369 & 0.0393 \\
N=8 & 0.1067 & 0.0439 \\
\hline
\end{tabular}
\label{tab:texture_index_comparison_uniaxial}
\end{table}
\newpage

\subsubsection{Local stress distribution} 
\begin{figure}[htbp]
\centering
\includegraphics[width=1\textwidth]{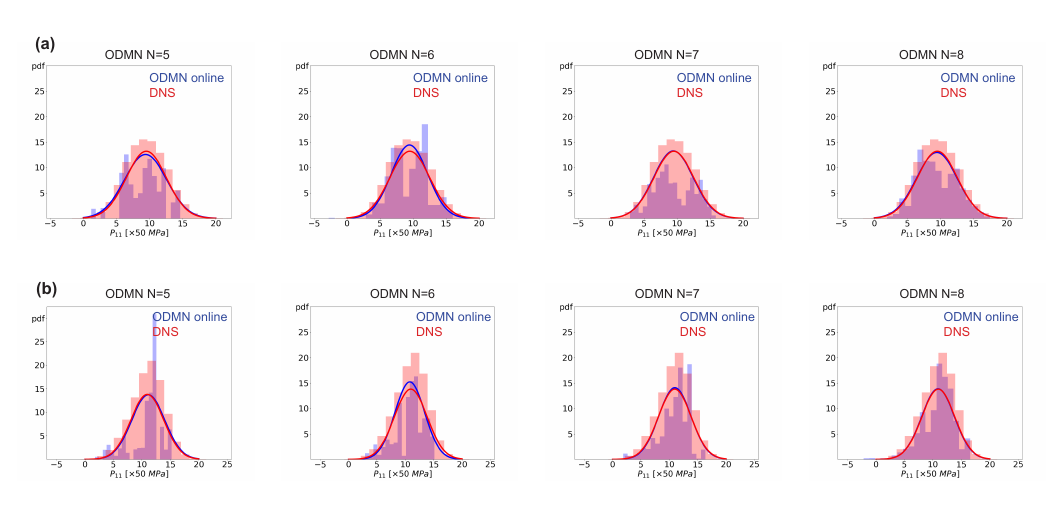}
\caption{
    Comparison of stress distributions predicted by ODMN model and DNS under uniaxial tensile loading at $F_{11}=1.3$ with a strain rate of $\dot{F}_{11}=1$ for (a) random textured and (b) preferred orientation RVEs.
}
\label{fig: stress_dist_singlePhase}
\end{figure}

To evaluate the ODMN model's capability to predict local stress distributions, its predictions were compared with those from DNS under uniaxial tensile loading at $F_{11} = 1.3$. As shown in Fig.~\ref{fig: stress_dist_singlePhase}, the stress distributions predicted by the ODMN model demonstrate strong agreement with the DNS results, particularly in terms of overall distribution patterns and trends. This indicates that the ODMN model effectively captures the local stress distribution across the RVE, contributing to accurate predictions of the homogenized mechanical response.

\newpage

\subsection{Two-phase polycrystal}
The two-phase polycrystalline RVE employed in this study was discretized into a 45x45x45 voxel of 100 grains. Among these, 30 grains belong to Phase 1, which exhibits a nearly random crystallographic texture, while the remaining 70 grains correspond to Phase 2, characterized by a single crystal orientation. The volume fraction of Phase 1 is 32.45\%, reflecting the relative distribution of the two constituent phases.

\begin{figure}[htbp]
\centering
\includegraphics[width=1.0\textwidth]{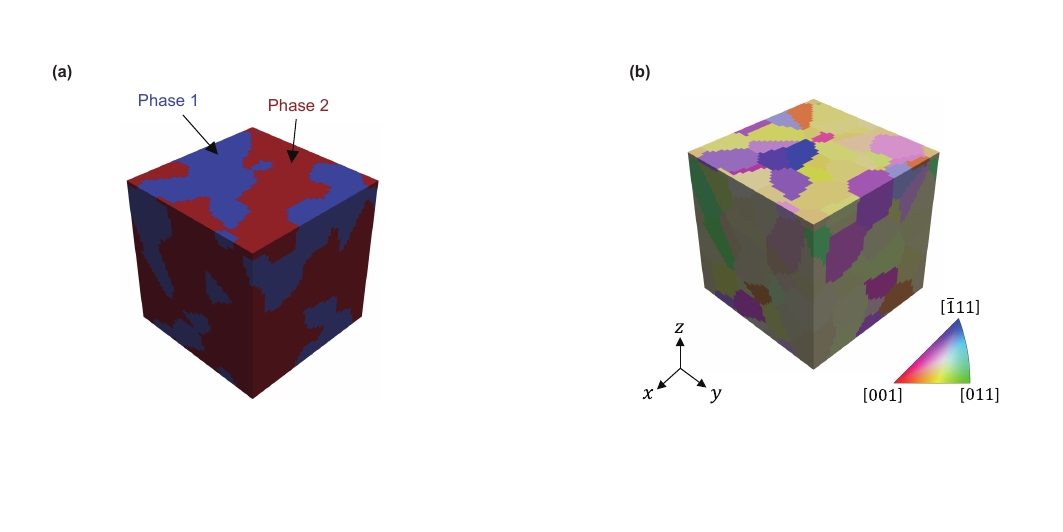}
\caption{
    Two-phase polycrystalline RVE was used to train the ODMN model. (a) Phase distribution: Phase 1 (blue) consists of nearly random texture, while Phase 2 (red) exhibits a single crystal orientation. (b) Crystallographic orientation map, where the color map represents different orientations as indicated in the legend.
}
\label{fig:2phase_RVE_geometry}
\end{figure}

The phase distribution of the RVE is shown in Fig.~\ref{fig:2phase_RVE_geometry}, where Phases 1 and 2 are visualized in blue and red, respectively. The crystallographic orientations of individual grains are further illustrated in the accompanying orientation map, with colors indicating distinct crystallographic orientations as defined by the legend.

\newpage

\subsubsection{Offline training results}
\begin{figure}[htbp]
\centering
\includegraphics[width=1\textwidth]{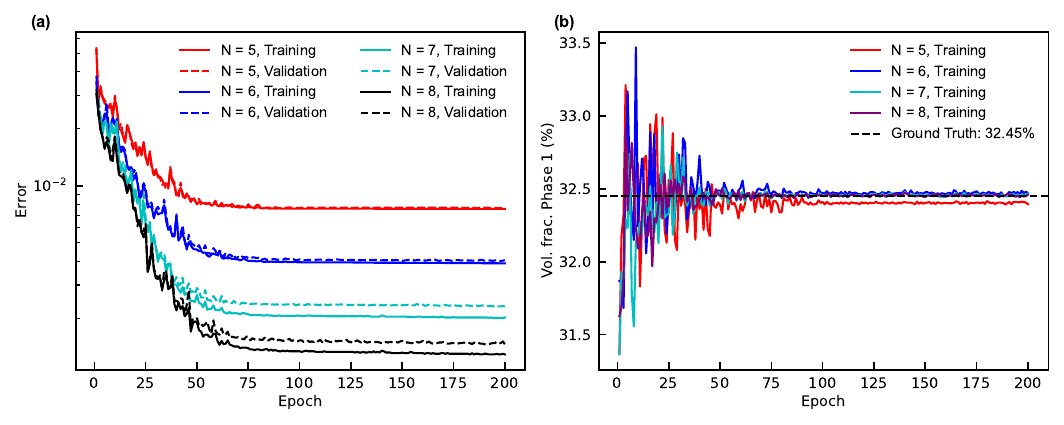}
\caption{
    (a) Training curve and (b) predicted volume fraction of Phase 1 over training epochs during the offline training of the ODMN model applied to the two-phase polycrystalline RVE.
}
\label{fig:training_curve_dualPhase}
\end{figure}

The performance of the ODMN model for the two-phase polycrystalline RVE was evaluated through offline training with varying hierarchical levels $N$. As \( N \) increased, the model's learning capability improved, evidenced by reduced training errors. Fig.~\ref{fig:training_curve_dualPhase}(a) presents the training curves, showing that higher values of \( N \) lead to decreased errors, reaching as low as 0.1\%. This demonstrates the model's effectiveness in capturing the complex microstructural characteristics between the two constituent phases within the RVE.

A notable feature of the two-phase RVE is the ODMN model's ability to learn the volume fractions of each phase during training. Fig.~\ref{fig:training_curve_dualPhase}(b) illustrates the predicted volume fraction of Phase 1 as the training progresses over epochs, which converges and stabilizes around the actual value of 32.45\%. This convergence demonstrates that the ODMN model accurately captures the phase distribution within the two-phase RVE. By simultaneously learning phase volume fractions and crystallographic orientations, the model effectively represents the microstructural characteristics of the two-phase polycrystal RVE.

\newpage

\begin{figure}[htbp]
\centering
\includegraphics[width=1.0\textwidth]{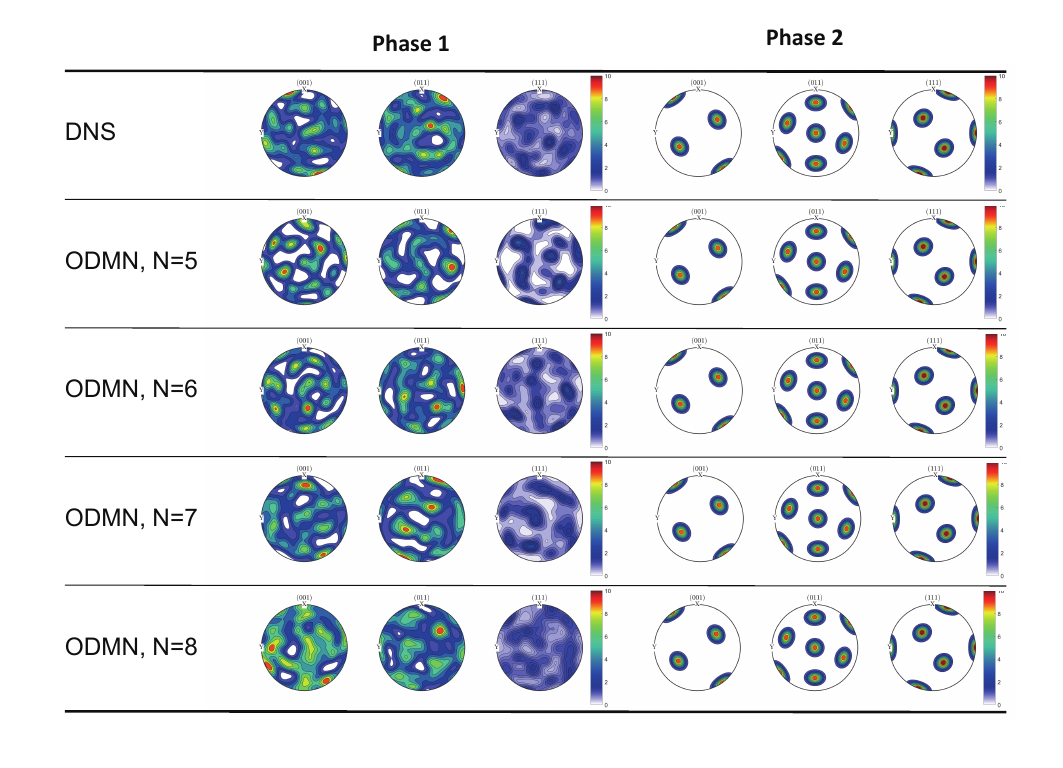}
\caption{
    Pole figures for the two-phase RVE, comparing DNS and the trained ODMN model.
}
\label{fig:poleFigure_2phase}
\end{figure}

Similar to the single-phase RVE, the trainable parameters of the ODMN model can be visualized using pole figures to evaluate its capability in capturing crystallographic orientation distributions. Fig.~\ref{fig:poleFigure_2phase} compares pole figures from the ODMN model with those from DNS results. As the number of hierarchical layers $N$ increases, the pole figures demonstrate closer alignment with the DNS results, emphasizing the ODMN model's improved accuracy in predicting crystallographic textures.

The results in Table~\ref{tab:texture_index_comparison_2phase} further support these observations. As model complexity increases, the DODF (\( \hat{T}^d \)) for both phases decreases, indicating the ODMN model's effectiveness in capturing texture distributions. Furthermore, the predicted volume fraction of Phase 1 converges closely to the actual value of 32.45\% as \( N \) increases. These findings highlight the model's ability to accurately represent both texture and phase distribution in the two-phase RVE with increasing hierarchical levels.

\begin{table}[htbp]
\centering
\caption{Normalized texture index of the difference ODF ($\hat{T}^d$) and predicted volume fraction of Phase 1 for the trained ODMN models with varying hierarchical levels \( N \). }
\begin{tabular}{lccc}
\hline
ODMN & Vol. frac. (Phase 1) & $\hat{T}^d$(Phase 1) & $\hat{T}^d$(Phase 2)\\
\hline
N = 5 & 32.50\% & 1.5208 & 0.00043 \\
N = 6 & 32.50\% & 0.9328 & 0.00042 \\
N = 7 & 32.47\% & 0.6605 & 0.00012 \\
N = 8 & 32.45\% & 0.5575 & 0.00006 \\
\hline
\end{tabular}
\label{tab:texture_index_comparison_2phase}
\end{table}

\subsubsection{Stress-strain predictions}

During the online prediction phase, the two-phase RVE was simulated using a phenomenological crystal plasticity model~\cite{roters2019damask}. Distinct material properties were assigned to each phase, with elastic behavior governed by generalized Hooke's law (see~\ref{appendixF}) and plastic behavior governed by the crystal plasticity model (see~\ref{appendixE}). The material parameters for Phase 1 are shown in Table~\ref{tab:crystal_plasticity_parameters_Phase1} and for Phase 2 in Table~\ref{tab:crystal_plasticity_parameters_Phase2}. These material properties were utilized to evaluate the ODMN model's capability to predict the mechanical behavior of the multi-phase polycrystal system under various loading conditions.

\begin{table}[htbp]
\centering
\caption{Phenomenological crystal plasticity parameters for Phase1}
\label{tab:crystal_plasticity_parameters_Phase1}
\renewcommand{\arraystretch}{1.2} 
\setlength{\tabcolsep}{10pt} 
\begin{tabular}{cccccccc}
\hline
\textbf{$N_s$} & \textbf{$h_0^{\text{sl-sl}}$}(GPa) & \textbf{$\xi^{\alpha}_{\infty}$}(MPa) & \textbf{$\xi^{0}$}(MPa) & \textbf{$n$} & \textbf{$a$} & \textbf{$\dot{\gamma}_0$}$(\text{s}^{-1})$ & \textbf{$h_{\text{int}}^{\alpha}$} \\
\hline
12 & 1.02 & 266 & 76 & 20 & 3.7 & 0.001 & 0 \\
\hline
\end{tabular}

\vspace{0.5cm}

\begin{tabular}{cccc}
\hline
{$C_{11}$ (GPa)} & {$C_{12}$ (GPa)} & {$C_{44}$ (GPa)} & {$h^{\text{sl-sl}}$} \\
\hline
191 & 162 & 42.2 & [1, 1, 5.123, 0.574, 1.123, 1.123, 1] \\
\hline
\end{tabular}
\end{table}


\begin{table}[htbp]
\centering
\caption{Phenomenological crystal plasticity parameters for Phase2}
\label{tab:crystal_plasticity_parameters_Phase2}
\renewcommand{\arraystretch}{1.2} 
\setlength{\tabcolsep}{10pt} 
\begin{tabular}{cccccccc}
\hline
\textbf{$N_s$} & \textbf{$h_0^{\text{sl-sl}}$}(GPa) & \textbf{$\xi^{\alpha}_{\infty}$}(MPa) & \textbf{$\xi^{0}$}(MPa) & \textbf{$n$} & \textbf{$a$} & \textbf{$\dot{\gamma}_0$}$(\text{s}^{-1})$ & \textbf{$h_{\text{int}}^{\alpha}$} \\
\hline
12 & 1.02 & 88.6 & 25.3 & 20 & 3.7 & 0.001 & 0 \\
\hline
\end{tabular}

\vspace{0.5cm}

\begin{tabular}{cccc}
\hline
{$C_{11}$ (GPa)} & {$C_{12}$ (GPa)} & {$C_{44}$ (GPa)} & {$h^{\text{sl-sl}}$} \\
\hline
191 & 162 & 42.2 & [1, 1, 5.123, 0.574, 1.123, 1.123, 1] \\
\hline
\end{tabular}
\end{table}

The ODMN model was evaluated under cyclic and shear loading conditions to assess its predictive capability across different loading scenarios. For cyclic loading, a strain-controlled cycle with a maximum strain of $F_{11} = 1.2$ was applied at a strain rate of $\dot{F}_{11} = 1$. For shear loading, the RVE was subjected to simple shear up to $F_{21} = 0.3$ at a strain rate of $\dot{F}_{21} = 1$.

\begin{figure}[htbp]
\centering
\includegraphics[width=1\textwidth]{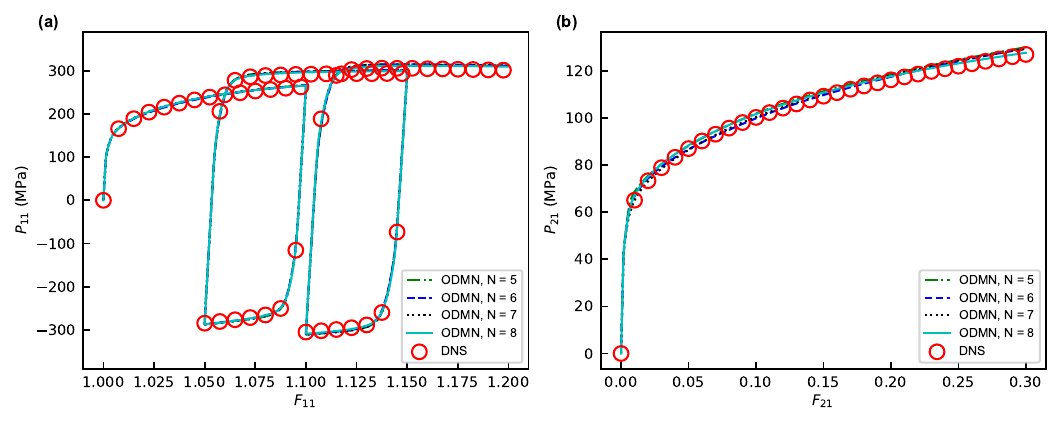}
\caption{
    Comparison of stress-strain curves between the ODMN model and DNS for the two-phase RVE under (a) cyclic and (b) simple shear loading.
}
\label{fig: SS_curve_dualPhase}
\end{figure}

As shown in Fig. \ref{fig: SS_curve_dualPhase}, the ODMN model's performance improves as the hierarchical level $N$ increases, with results closely matching DNS when $N \geq 7$. Beyond this value, further increases in $N$ yield diminishing returns in accuracy, indicating an optimal balance between model complexity and performance.


\subsubsection{Local stress distribution} 
\begin{figure}[htbp]
\centering
\includegraphics[width=1\textwidth]{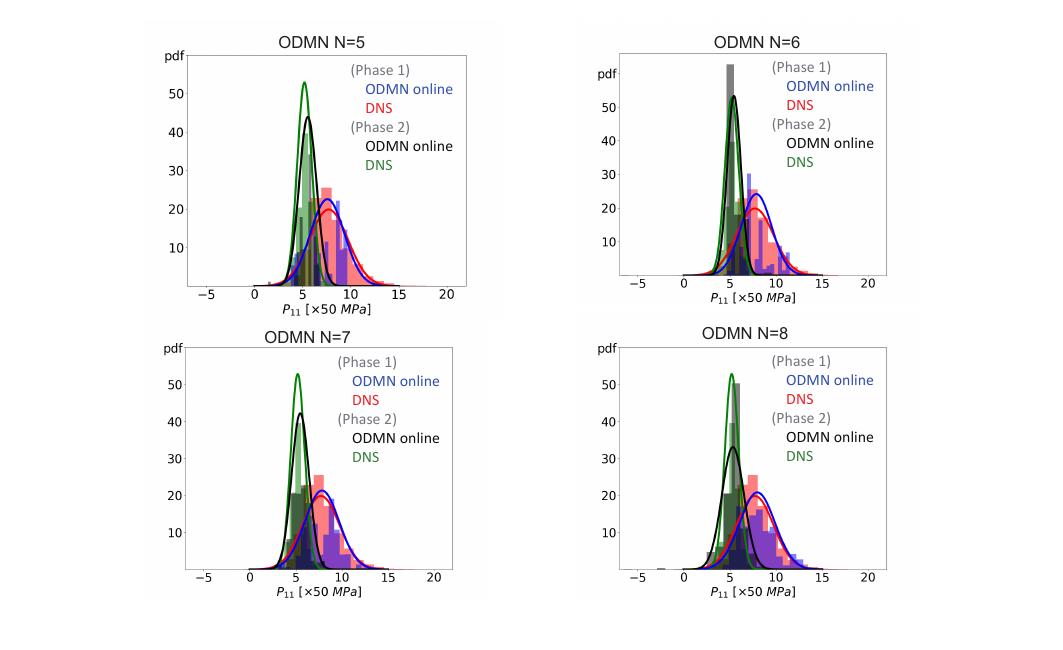}
\caption{
    Comparison of local stress distributions predicted by the ODMN model and DNS for the two-phase RVE under uniaxial tensile loading at $F_{11}=1.3$.
}
\label{fig:stress_dist_dualPhase}
\end{figure}

Local stress distributions within a two-phase RVE predicted from the ODMN model were compared with DNS results under uniaxial tensile loading at $F_{11} = 1.3$. As shown in Fig.~\ref{fig:stress_dist_dualPhase} and summarized in Table~\ref{tab:comparison_stress}, the ODMN model demonstrates strong agreement with DNS results, particularly in the overall trends of local stress distributions across both phases.

\begin{table}[htbp]
\centering
\caption{Comparison of local stress distributions between the ODMN model and DNS in the two-phase RVE.}
\begin{tabular}{lll}
\hline
 & {Phase 1}  & {Phase 2}  \\
  &   [MPa] &  [MPa]  \\
\hline
DNS         & 385.93 $\pm$ 100.24 & 260.06 $\pm$ 37.72     \\
ODMN, N=5     & 379.11 $\pm$ 99.30  & 277.57 $\pm$ 69.99    \\
ODMN, N=6     & 393.59 $\pm$ 82.48  & 272.43 $\pm$ 37.40    \\
ODMN, N=7     & 391.54  $\pm$ 93.44  & 273.87 $\pm$ 48.40   \\
ODMN, N=8     & 398.85 $\pm$ 95.74 & 266.43 $\pm$ 60.22     \\
\hline
\end{tabular}
\label{tab:comparison_stress}
\end{table}

Table~\ref{tab:comparison_stress} shows that the ODMN model effectively reproduces the stress distribution trends observed in DNS, particularly for Phase 1. Although some deviations are observed in Phase 2, the mean stress values align closely with DNS results, demonstrating the model's capability to accurately capture the overall local stress behavior in both phases.

\subsection{Computational cost}

The computational cost of the ODMN model was assessed by measuring the CPU time required for generating the training dataset, performing offline training, and conducting online predictions. Offline training computations were executed on an Apple M3 Max chip, while the online prediction phase utilized Intel\textsuperscript{®} Xeon\textsuperscript{®} Platinum 8480+2.0GHz CPUs.

The offline training involved optimizing model parameters over 200 epochs for various model architectures. The training times for random textured, preferred orientation, two-phase RVEs, and online prediction times under different loading conditions are summarized in Table~\ref{tab:computation_time_training_and_prediction}.



\begin{table}[htbp]
\centering
\caption{Wall time (in CPU-seconds) required for offline training and online predictions of the ODMN model for random textured, preferred orientation, and two-phase RVEs. $T_{\textit{offline}}$ denotes the total wall time for training, while $N_{\mathcal{F}}$ represents the number of fitting parameters in each model configuration. $T_{\textit{Tensile}}$, $T_{\textit{Cyclic}}$, and $T_{\textit{Shear}}$ correspond to the wall time for uniaxial tensile, cyclic, and shear loading conditions, respectively. }

\begin{tabular}{llllll}
\hline
\multicolumn{6}{c}{\textbf{Random Textured RVE}} \\
 & $T_{\textit{offline}}$ & $N_{\mathcal{F}}$ & $T_{\textit{Tensile}}$ & $T_{\textit{Cyclic}}$ & $T_{\textit{Shear}}$\\
 \hline
DNS & - & -              & 698880   &  885472  & 609056 \\
ODMN N=5 & 412   & 190  & 705      &  1129    & 806 \\
ODMN N=6 & 864   & 382  & 1186     &  2015    & 1432 \\
ODMN N=7 & 1858  & 766  & 2753     &  4208    & 3151 \\
ODMN N=8 & 4325  & 1534 & 7306     &  10683   & 8207 \\
\hline
\multicolumn{6}{c}{\textbf{Preferred Orientation RVE}} \\
 & $T_{\textit{offline}}$ & $N_{\mathcal{F}}$ & $T_{\textit{Tensile}}$ & $T_{\textit{Cyclic}}$ & $T_{\textit{Shear}}$\\
  \hline
DNS      & -      & -      & 774032    & 919408   & 610512      \\
ODMN N=5 & 362    & 190    & 756       & 1122     & 662      \\
ODMN N=6 & 812    & 382    & 1333      & 1968     & 1145      \\
ODMN N=7 & 1794   & 766    & 3054      & 3957     & 2379      \\
ODMN N=8 & 4524   & 1534   & 7879      & 10783    & 5827   \\
\hline  
\multicolumn{6}{c}{\textbf{two-phase RVE}} \\
 & $T_{\textit{offline}}$ & $N_{\mathcal{F}}$ &   & $T_{\textit{Cyclic}}$ & $T_{\textit{Shear}}$\\
  \hline
DNS & - & -              &    & 1272096  & 511840  \\
ODMN N=5 & 417    & 190  &    & 1766     & 760  \\
ODMN N=6 & 908    & 382  &    & 3423     & 1721  \\
ODMN N=7 & 2072    & 766  &   & 8065     & 3701  \\
ODMN N=8 & 5126    & 1534 &   & 18163    & 8207  \\
\hline\\
\end{tabular}
\label{tab:computation_time_training_and_prediction}
\end{table}

To quantitatively evaluate the performance improvement of the ODMN model compared to DNS, the speedup, $S_p$ is defined as:
\begin{equation}
S_p = \frac{T_{\text{DNS}}}{T_{\text{ODMN}}}
\end{equation}
where $T_{\text{DNS}}$ is the wall time required for DNS and $T_{\text{ODMN}}$ is the wall time required for ODMN simulations.

\begin{figure}[htbp]
\centering
\includegraphics[width=0.5\textwidth]{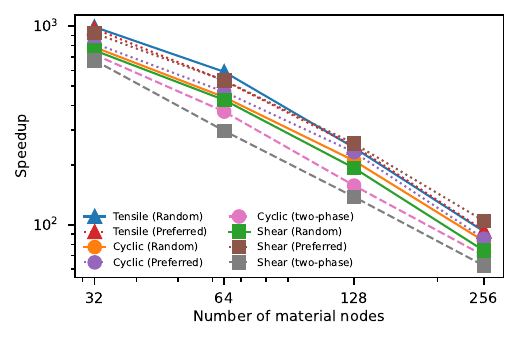}
\caption{
    Speedup ($S_p$) of the ODMN model relative to DNS as a function of the number of material nodes.
}
\label{fig: speedup}
\end{figure}

As illustrated in Fig.~\ref{fig: speedup}, the ODMN model achieves a speedup of two to three orders of magnitude compared to DNS despite being implemented in a Python-based in-house code without specific performance optimizations. Notably, with $N = 7$, the ODMN model strikes an optimal balance between accuracy and computational cost, demonstrating excellent performance in texture prediction and as a surrogate model. The speedup decreases linearly with increasing material nodes, consistent with trends in IMN~\cite{noels2022micromechanics}. The DNS was performed using the DAMASK FFT solver, further underscoring the significant computational advantages of the ODMN model, even without hardware-specific tuning.

\section{Conclusions} \label{sec06}

This study introduces the ODMN as an innovative surrogate model for polycrystalline materials. The ODMN employs a partitioned RVE architecture with two core mechanisms: (1) an interaction mechanism that enforces stress-equilibrium across subregions to capture internal microstructural mechanics, and (2) an orientation-aware mechanism, which learns and represents crystallographic orientations, enabling accurate texture distribution predictions. By integrating these mechanisms, the ODMN effectively encodes key microstructural features, such as phase distributions and crystallographic textures, providing a robust and computationally efficient framework for mechanistic learning.

The performance and versatility of the ODMN framework were demonstrated through two comprehensive case studies. The first focused on single-phase polycrystalline RVEs with random textures and preferred orientations. The ODMN successfully predicted stress-strain behavior and texture evolution under diverse loading conditions, including uniaxial tensile loading at varying strain rates, cyclic loading, and simple shear. These results highlight the model’s capability to generalize across different deformation paths. The second case extended the ODMN to two-phase polycrystalline RVEs, accurately capturing distinct phase distributions and crystallographic textures. The model exhibited excellent agreement with DNS results, accurately predicting mechanical responses across complex, multi-phase material systems, underscoring its adaptability and generalizability.

The advantages of the ODMN framework can be summarized as follows:
\begin{enumerate}
    \item Computational efficiency: Achieves significant computational speed-ups compared to traditional full-field models while maintaining high accuracy.

    \item Versatility: Demonstrates adaptability to diverse microstructural configurations, effectively modeling both single-phase and two-phase polycrystal RVE under various loading conditions.

    \item Predictive accuracy: Provides precise predictions of mechanical behavior and texture evolution, critical for understanding material responses under complex deformation scenarios.
\end{enumerate}

Future work will focus on expanding the ODMN framework in several directions. A primary objective is extending its application to two-scale simulations, facilitating the modeling of industrial-scale components, particularly for processes such as metal forming. Another promising direction involves incorporating damage mechanics into the framework, enabling predictions of material degradation and failure, thereby offering a more comprehensive tool for simulating real-world material performance.

In summary, the ODMN framework bridges the gap between microstructural features and macroscopic material responses, establishing itself as a powerful, efficient, and versatile tool for advancing computational materials science. Its capability to accurately represent complex microstructural geometries and predict mechanical responses marks a significant step forward in multiscale material modeling and design.

\section*{Acknowledgements}
This work is supported by the National Science
and Technology Council, Taiwan, under Grant 111-2221-E-002-054-
MY3 and 112-2221-E-007-028. We are grateful for the computational
resources and support from the NTUCE-NCREE Joint Artificial Intelligence Research Center and the National Center of High-performance Computing (NCHC). In addition, we sincerely thank the anonymous reviewers for their constructive comments.

\bibliographystyle{unsrt}  
\bibliography{references}

\appendix
\renewcommand{\thesection}{Appendix \Alph{section}}  
\renewcommand{\theequation}{\Alph{section}.\arabic{equation}}  
\setcounter{equation}{0}  
\renewcommand{\thesubsection}{\Alph{section}.\arabic{subsection}}  

\section{Rotation matrices in orientation-aware mechanisms}
\label{appendixA}

Each material node $\mathcal{M}^i$ is characterized by a unique set of orientation angles $\alpha^i, \beta^i, \gamma^i$, which correspond to sequential rotations about the x, y, and z axes, respectively. The rotation matrices associated with the Cauchy stress in Eq. \eqref{eq:R1}
 are expressed as follows:

\begin{equation}
    \mathbf{X^{R_1}}_{(1,1)}= 1, \quad
\mathbf{X^{R_1}}_{([2,3,4],[2,3,4])}= \textbf{R}^{\text{in},\sigma}(\alpha^i) \quad
\mathbf{X^{R_1}}_{([5,6],[5,6])}= \textbf{R}^{\text{out}}(\alpha^i)
\end{equation}

\begin{equation}
    \mathbf{Y^{R_1}}_{(2,2)}= 1, \quad
\mathbf{Y^{R_1}}_{([1,3,5],[1,3,5])}= \textbf{R}^{\text{in},\sigma}(-\beta^i) \quad
\mathbf{Y^{R_1}}_{([4,6],[4,6])}= \textbf{R}^{\text{out}}(-\beta^i)
\end{equation}

\begin{equation}
    \mathbf{Z^{R_1}}_{(3,3)}= 1, \quad
\mathbf{Z^{R_1}}_{([1,2,6],[1,2,6])}= \textbf{R}^{\text{in},\sigma}(\gamma^i) \quad
\mathbf{Z^{R_1}}_{([4,5],[4,5])}= \textbf{R}^{\text{out}}(\gamma^i)
\end{equation}

Similarly, the rotation matrices corresponding to the infinitesimal strain in Eq. \eqref{eq:R2}
 are given by:

\begin{equation}
    \mathbf{X^{R_2}}_{(1,1)}= 1, \quad
\mathbf{X^{R_2}}_{([2,3,4],[2,3,4])}= \textbf{R}^{\text{in},\epsilon}(\alpha^i) \quad
\mathbf{X^{R_2}}_{([5,6],[5,6])}= \textbf{R}^{\text{out}}(\alpha^i)
\end{equation}

\begin{equation}
    \mathbf{Y^{R_2}}_{(2,2)}= 1, \quad
\mathbf{Y^{R_2}}_{([1,3,5],[1,3,5])}= \textbf{R}^{\text{in},\epsilon}(-\beta^i) \quad
\mathbf{Y^{R_2}}_{([4,6],[4,6])}= \textbf{R}^{\text{out}}(-\beta^i)
\end{equation}

\begin{equation}
    \mathbf{Z^{R_2}}_{(3,3)}= 1, \quad
\mathbf{Z^{R_2}}_{([1,2,6],[1,2,6])}= \textbf{R}^{\text{in},\epsilon}(\gamma^i) \quad
\mathbf{Z^{R_2}}_{([4,5],[4,5])}= \textbf{R}^{\text{out}}(\gamma^i)
\end{equation}

The out-of-plane rotation matrix $\mathbf{R}^{\text{out}}$ for a rotation angle $\theta$ is defined as:
\begin{equation}
    \textbf{R}^{\text{out}}=\begin{bmatrix}
\cos\theta & -\sin\theta \\
\sin\theta & \cos\theta
\end{bmatrix}
\end{equation}

For in-plane rotations, the rotation matrices for stress, $\mathbf{R}^{\text{in},\sigma}$, and strain, $\mathbf{R}^{\text{in},\epsilon}$, with a rotation angle $\theta$ are expressed as:

\begin{equation}
    \textbf{R}^{\text{in},\sigma}(\theta)=\begin{bmatrix}
\cos^2\theta & \sin^2\theta & 2\sin\theta\cos\theta \\
\sin^2\theta & \cos^2\theta & -2\sin\theta\cos\theta \\
-\sin\theta\cos\theta & \sin\theta\cos\theta & \cos^2\theta-\sin^2\theta
\end{bmatrix}
\end{equation}

\begin{equation}
   \textbf{R}^{\text{in}, \epsilon}(\theta)=\begin{bmatrix}
\cos^2\theta & \sin^2\theta & \sin\theta\cos\theta \\
\sin^2\theta & \cos^2\theta & -\sin\theta\cos\theta \\
-2\sin\theta\cos\theta & 2\sin\theta\cos\theta & \cos^2\theta-\sin^2\theta
\end{bmatrix}
\end{equation}

\section{Binary homogenization operator}
\label{appendixB}

During the offline training stage, the stiffness matrix is homogenized using the binary homogenization operator $\mathbb{H}_2$, as described in Eq. \eqref{eq:recursive_homogenization}. The homogenized stiffness is computed through the binary homogenization operator as follows:

\begin{equation}\label{eq:H2 function}
\bar{\mathbf{C}} = \mathbb{H}_2(\mathbb{C}^0, \mathbb{C}^1, \mathcal{K}^0, \mathcal{K}^1, \vec{\mathbf{N}}) = f^0 \mathbb{C}^0 + f^1 \mathbb{C}^1 - f^0 f^1 (\mathbb{C}^0 - \mathbb{C}^1)
\mathbf{Q}(\mathbb{C}^0 - \mathbb{C}^1)
\end{equation}

Here, $f^0$ and $f^1$ represent the volume fractions of the left and right sub-trees, respectively, and are computed as follows:

\begin{equation}
f^0 = \sum_{i \in \mathcal{K}^0} W^i \bigg/ \sum_{i \in \mathcal{K}^0 \cup \mathcal{K}^1} W^i
\end{equation}

\begin{equation}
f^1 = \sum_{i \in \mathcal{K}^1} W^i \bigg/ \sum_{i \in \mathcal{K}^0 \cup \mathcal{K}^1} W^i
\end{equation}

The matrix $\mathbf{Q}$ is computed as follows:

\begin{equation}\label{eq: Q}
\mathbf{Q} = \mathbf{H}\mathbf{S}^{-1}\mathbf{H}^T, \quad
\mathbf{S} = \mathbf{H}^T(f^1\bar{\mathbb{C}}^0 + f^0\bar{\mathbb{C}}^1)\mathbf{H}
\end{equation}

where the matrix $\mathbf{H}(\vec{\mathbf{N}})$ is defined as:

\begin{equation}
    \textbf{H}(\vec{\mathbf{N}} )=\begin{bmatrix}
N_0 & 0 & 0 \\
0 & N_1 & 0 \\
0 & 0 & N_2 \\
N_1 & N_0 & 0 \\
N_2 & 0 & N_0 \\
0 & N_2 & N_1
\end{bmatrix}
\end{equation}

Here, the vector $\vec{\mathbf{N}}$ is given by:

\begin{equation}
    \vec{\mathbf{N}} = \begin{bmatrix}
N_0 \\
N_1 \\
N_2
\end{bmatrix} = \begin{bmatrix}
\cos(2\pi\phi)\sin(\pi \theta) \\
\sin(2\pi\phi)\sin(\pi \theta) \\
\cos(\pi \theta)
\end{bmatrix}
\end{equation}

\section{Residual calculation of the Hill-Mandel condition}
\label{appendixD}

The system residual $\mathbf{r}$, which quantifies the degree to which the Hill–Mandel condition is violated, is given by:

\begin{equation} 
    \mathbf{r} = \sum_{i=0}^{2^N-1} W^i (\mathbf{D}^i)^T \text{vec}(\mathbf{P}^i)
\end{equation}

Here, $\text{vec}(\mathbf{P}^i)$ represents the vectorized form of the first Piola-Kirchhoff stress tensor at node $\mathcal{M}^i$, defined as:

\begin{equation}
    \text{vec}(\mathbf{P}^i) = [P_{11}, P_{21}, P_{31}, P_{12}, P_{22}, P_{32}, P_{13}, P_{23}, P_{33}]
\end{equation}

The matrix $\mathbf{D}^i$ is composed of the interaction variables $\alpha^{i,j}$ and the force-balance direction $\mathbf{N}^j$, and is defined as:

\begin{equation}
    \mathbf{D}^i = \left[\alpha^{i,j} \mathbf{R}^j \quad \text{for } j = 0,\dots,2^N-2 \right]
\end{equation}

The force-balance direction vector $\mathbf{N}^j$ is expressed as:

\begin{equation}
    \mathbf{R}^j = \begin{bmatrix} 
    N_0\\ 
    N_1\\ 
    N_2\\ 
    \end{bmatrix}
\end{equation}

The matrix $\mathbf{R}^j$, which maps the components of $\mathbf{N}^j$, is defined as:

\begin{equation}
    \mathbf{R}^j = \begin{bmatrix}
    N_0 & 0 & 0\\ 
    0 &  N_0 & 0 \\ 
    0 & 0 & N_0\\ 
    N_1 & 0 & 0 \\ 
    0 &  N_1 & 0 \\ 
    0 & 0 & N_1\\ 
    N_2 & 0 & 0 \\ 
    0 &  N_2 & 0 \\ 
    0 & 0 & N_2
    \end{bmatrix}
\end{equation}

To iteratively adjust the solution and minimize the residual, the assembly matrix $\mathbf{A}$, which collects the interaction variables $\mathbf{a}^j$ for each interaction mechanism, is updated. The assembly matrix is defined as:

\begin{equation}
    \mathbf{A} = \begin{bmatrix} (\mathbf{a}^0)^T, \ \dots ,\ (\mathbf{a}^{2^N-2})^T \end{bmatrix}
\end{equation}

The matrix update is performed using the residual and its gradient with respect to $\mathbf{A}$:

\begin{equation}\label{eq: deltaA}
    \delta \mathbf{A} = -\frac{\partial \mathbf{r}}{\partial \mathbf{A}} \mathbf{r}
\end{equation}

\begin{equation}\label{eq: correctA}
    \mathbf{A} \leftarrow \mathbf{A} + \delta \mathbf{A}
\end{equation}

This iterative process continues until the residual $\mathbf{r}$ is sufficiently small, ensuring that the system reaches equilibrium and satisfies the Hill–Mandel condition.

\section{Phenomenological Crystal Plasticity Model}
\label{appendixE}

In this study, the phenomenological crystal plasticity model is used to describe the material behavior as the local material law. This model is adapted from DAMASK and does not account for deformation twinning \cite{roters2019damask}. The plastic velocity gradient $\mathbf{L}_p$ is computed as the cumulative contribution of dislocation slip across all slip systems $\alpha$, as described by the following equation:

\begin{equation}
    \mathbf{L}_p = \sum_{\alpha} \dot{\gamma}^\alpha (\mathbf{s}^\alpha_s \otimes \mathbf{n}^\alpha_s)
\end{equation}

Here, $\mathbf{s}^\alpha_s$ is the unit vector in the slip direction, $\mathbf{n}^\alpha_s$ is the unit vector normal to the slip plane, and $\dot{\gamma}^\alpha$ is the shear rate for slip system $\alpha$.

The evolution of slip resistance $\xi^\alpha$ is governed by a dynamic equation that defines its transition from the initial value $\xi_0^\alpha$ to the saturation value $\xi_{\infty}^{\alpha}$. The following equation describes this evolution:

\begin{equation}
    \dot{\xi}^\alpha = h_0^{\text{s-s}}(1+h_{\text{int}}^\alpha) \times\sum_{\alpha'}^{N_s}\left | \dot{\gamma}^{\alpha'} \right |{\left | 1-\frac{\xi ^{\alpha'}}{\xi ^{\alpha'}_\infty} \right |}^{a}\text{sgn}\left(1-\frac{\xi ^{\alpha'}}{\xi ^{\alpha'}_\infty}\right)h^{\alpha\alpha'}
\end{equation}

The shear rate $\dot{\gamma}^\alpha$ for each slip system is computed as a function of slip resistance $\xi^\alpha$ and resolved shear stress $\tau^\alpha$:

\begin{equation}
    \dot{\gamma}^\alpha = \dot{\gamma}^\alpha_0 {\left | \frac{\tau^\alpha}{\xi^\alpha} \right |}^n \text{sgn}(\tau^\alpha)
\end{equation}

The resolved shear stress $\tau^\alpha$ is calculated using Schmid's law and the Mandel stress $\mathbf{M}^p$:

\begin{equation}
    \tau^\alpha = \mathbf{M}^p \cdot (\mathbf{s}^\alpha \otimes \mathbf{n}^\alpha)
\end{equation}


\section{Generalized Hooke’s law}
\label{appendixF}

In this study, the Generalized Hooke’s Law is employed to describe the elastic behavior of materials as the local material law. The generalized form of Hooke’s law relates the second Piola–Kirchhoff stress tensor, $\mathbf{S}$, to the Green-Lagrange strain tensor, $\mathbf{E}$, via the material stiffness tensor, $\mathbb{C}$, as follows:

\begin{equation}
    \mathbf{S} = \mathbb{C} : \mathbf{E}
\end{equation}

In this equation:
\begin{itemize}
    \item $\mathbf{S}$ represents the second Piola–Kirchhoff stress tensor, a stress measure in the material's reference configuration.

    \item $\mathbb{C}$ is the fourth-order material stiffness tensor, which encapsulates the material’s elastic properties.

    \item $\mathbf{E}$ is the Green-Lagrange strain tensor, describing the strain experienced by the material during deformation. The operator $:$ denotes the double contraction between tensors.
\end{itemize}

The Green-Lagrange strain tensor $\mathbf{E}$ is defined directly in terms of the elastic deformation gradient, $\mathbf{F}^e$, as:

\begin{equation}
    \mathbf{E} = \frac{1}{2} \left( (\mathbf{F}^e)^{T} \mathbf{F}^e - \mathbb{I} \right)
\end{equation}
Where:
\begin{itemize}
    \item $\mathbf{F}^e$ is the elastic deformation gradient, representing the reversible part of the deformation.
    \item $\mathbb{I}$ is the second-order identity tensor.
\end{itemize}

\end{document}